\documentclass[12pt,a4paper]{article}
\usepackage{amsmath,amssymb,graphicx,epsfig,cite,color}
\usepackage{makecell}
\usepackage[left=2.7cm,right=2.7cm,top=2.5cm,bottom=2.5cm]{geometry}
\usepackage{subfigure}
\usepackage[T1]{fontenc}
	\usepackage{hyperref}
	\usepackage{float}   
	\usepackage{array} 
	\usepackage{makecell}
	\usepackage{multirow}
	\usepackage[table]{xcolor}
	\usepackage{booktabs}
	\usepackage{array}
	\usepackage{colortbl}
	\usepackage{orcidlink}
	
	\definecolor{tabcolor1}{rgb}{.105,.410,.113}
	\definecolor{tabcolor2}{rgb}{.425,.130,.303}
	\definecolor{tabcolor3}{rgb}{1.00,.0,.0}
	
	\usepackage{xcolor,colortbl}

	\usepackage{graphicx}
	\usepackage{lscape}
	\usepackage{appendix}
	\usepackage{multirow}
	\definecolor{DarkPastelGreen}{rgb}{0.01,0.69,0.28}
	\definecolor{amethyst}{rgb}{0.6, 0.4, 0.8}
	\definecolor{darkcyan}{rgb}{0.0, 0.55, 0.55}
	\definecolor{lightcyan}{rgb}{0.88, 1.0, 1.0}
	\definecolor{antiquefuchsia}{rgb}{0.57, 0.36, 0.51}
	\definecolor{brightube}{rgb}{0.82, 0.62, 0.91}
	\definecolor{brilliantlavender}{rgb}{0.96, 0.73, 1.0}
	\definecolor{bazaar}{rgb}{0.6, 0.47, 0.48}
	\allowdisplaybreaks[1]
	
\begin{document}

		\title{
			{\Large
				{\bf
					\color{amethyst}{Quintic Modification to Lifshitz Quasi-topological Black Holes }
				}
			}
		}

		\author{\vspace{1cm}\\
			{\small
				A.~Bazrafshan	$^{1}$,
				A. R. Olamaei \orcidlink{0000-0003-3529-3002}
				\thanks{Corresponding author, E-mail: olamaei@gmail.com}
				$^{1,2}$,
				M. Ghanaatian $^{1}$	
			}
			\\
			{\small $^1$ Department of Physics, Jahrom University, Jahrom, P.~O.~Box 74137-66171, Iran}\\
			{\small $^2$  School of Physics, Institute for Research in Fundamental Sciences (IPM),}\\
			{\small  P. ~O. ~Box 19395-5531, Tehran, Iran}\\
		} 
		
		\date{}

		\begin{titlepage}
			\maketitle
			\thispagestyle{empty}
			
			\begin{abstract}
				We extend the analysis of Lifshitz black holes to quintic order in five-dimensional quasi-topological gravity coupled to a massive Abelian vector field. Starting from a static ansatz with a constant-curvature horizon, we derive the reduced field equations and identify the radially conserved quantity of the one-dimensional effective system. We then analyze the algebraic conditions that permit Lifshitz backgrounds, both in the absence and in the presence of the massive vector field. 
				Since closed-form black-hole solutions are not available for the generic
				quintic theory, we construct numerical solutions by integrating the
				field equations outward from regular near-horizon data.
				 We present solutions for the relativistic branch \(z=1\) and the Lifshitz branch \(z=2\), covering the three horizon topologies \(k=-1,0,+1\). 
				The numerical profiles remain qualitatively consistent with those
				found previously in cubic and quartic quasi-topological Lifshitz
				gravity. A controlled planar comparison through the genuine quartic
				point \(\hat\mu_5=0\) further shows that the very small quintic
				couplings used in the representative examples produce nearly quartic
				profiles, whereas larger admissible values of \(\hat\mu_5\) lead to
				clearly resolved changes in both the normalized radial profiles and
				the Hawking temperature.
				 We also compute the Wald entropy and Hawking temperature and examine the
				 thermal behavior through dimensionless logarithmic temperature--entropy
				 plots. For the representative branches considered here, the heat capacity
				 is positive. For a fixed-theory planar \(z=2\) family, the conserved radial
				 charge is further used to determine the energy density; the differential
				 and integrated first laws are verified numerically to the precision of the
				 computation, and the corresponding Helmholtz free-energy density is found
				 to be negative relative to the Lifshitz vacuum normalization adopted here.
			\end{abstract}

		\end{titlepage}
		\hypersetup{pageanchor=true}
		
		\section{Introduction}
		
		Gauge/gravity duality offers a powerful framework in which strongly coupled quantum systems can be studied via classical gravitational dynamics in one higher dimension \cite{Maldacena1998,Gubser1998,Witten1998}. Its most thoroughly understood realization is the AdS/CFT correspondence, but holographic techniques have also been generalized to systems whose scaling behavior departs from relativistic invariance \cite{Hartnoll2009,McGreevy2009}. A particularly prominent class of such nonrelativistic systems is characterized by Lifshitz scaling,
		\begin{equation}
			t \rightarrow \lambda^{z} t, \qquad
			x_i \rightarrow \lambda x_i ,
		\end{equation}
		where \(z\) denotes the dynamical critical exponent.
		For \(z=1\), the scaling reduces to the relativistic symmetry familiar
		from conformal field theories, whereas \(z\neq1\) encodes an anisotropic
		scaling between time and space. Gravitational geometries with Lifshitz
		asymptotics therefore provide a natural setting for holographic models
		of nonrelativistic critical phenomena and related condensed-matter
		systems
		\cite{Kachru2008,Taylor2008,Mann2009,BalasubramanianMcGreevy2009,
			DanielssonThorlacius2009,Taylor2016}.
		A broad class of Lifshitz black holes and black branes has subsequently
		been constructed in different matter-coupled and higher-curvature
		settings
		\cite{TarrioVandoren2011,AyonBeatoGarbarzGiribetHassaine2009,
			AlvarezAyonBeatoGonzalezHassaine2014,
			BravoGaeteJuarezAubryVelazquezRodriguez2022}.
		Their transport properties and thermodynamics have also been studied
		from several perspectives
		\cite{Pang2010,DeveciogluSarioglu2011}.
		
		From a gravitational standpoint, it is equally natural to incorporate higher-curvature corrections to the Einstein-Hilbert action. These corrections emerge in effective descriptions of gravity and extend the space of gravitational couplings accessible within holographic models \cite{deBoer2010,CamanhoEdelstein2010}. This is significant because pure Einstein gravity contains only a restricted set of parameters and therefore cannot capture the full range of possible dual field theories, especially when quantities such as central charges, transport coefficients, and thermodynamic response functions are permitted to vary independently \cite{MyersPaulosSinha2010,DehghaniVahidinia2013}.
		Lovelock gravity is particularly well suited to this context because its higher-curvature densities yield field equations that contain at most second derivatives of the metric \cite{Lovelock1971}. In a fixed low dimension, however, the higher-order Lovelock terms either vanish identically or become topological, which limits their usefulness for five-dimensional holographic applications.
		
		Quasi-topological gravity provides a means to preserve some of the appealing features of Lovelock theory while still yielding nontrivial higher-curvature dynamics in five dimensions \cite{MyersRobinson2010,OlivaRay2010}.
		Quasi-topological interactions are specifically designed so that the field equations simplify considerably on highly symmetric backgrounds---including static black-hole ans\"atze---despite the fact that the theory is not second order for a completely generic metric. This property makes quasi-topological gravity a particularly useful laboratory for investigating how additional curvature couplings modify black-hole solutions and their holographic interpretation \cite{MyersPaulosSinha2010,DehghaniVahidinia2013}.
		
		Asymptotically Lifshitz black holes in quasi-topological gravity have previously been investigated at cubic and quartic order \cite{BrennaDehghaniMann2011,GhanaatianBazrafshanBrenna2014}. The cubic theory showed that quasi-topological interactions can support Lifshitz black holes in five dimensions, and that the resulting solutions can be constructed numerically by combining near-horizon expansions with shooting methods.
		The quartic extension showed that this structure continues to hold when an additional higher-curvature coupling is introduced; the quartic term alters the black-hole profiles and thermodynamic behavior but preserves the overall pattern seen in the lower-order solutions.
		Quintic quasi-topological gravity, a higher-order extension of this framework, has been constructed more recently. Black-hole solutions in the quintic theory have begun to be investigated in a variety of other settings, including static higher-dimensional solutions, thermodynamic analyses, boundary terms, conserved quantities, and charged black-hole geometries \cite{CisternaGuajardoHassaineOliva2017,BazrafshanNaeimipourOlamaeiGhanaatian2019,BazrafshanOlamaei2020,OlamaeiBazrafshanGhanaatian2024}. These advances provide strong motivation to study Lifshitz black holes within the quintic theory.
		
		
		In this paper, we extend the Lifshitz black-hole analysis of the cubic
		and quartic quasi-topological theories to the representative quintic
		density adopted here. We emphasize that quintic order is not the highest
		curvature order available in five dimensions: quasi-topological and
		generalized quasi-topological densities have been constructed at
		arbitrarily high curvature order in \(D\geq5\)
		\cite{BuenoCanoHennigar2020,BuenoCanoHennigarLuMoreno2023}. This extension is not merely a formal exercise; it introduces an additional coupling that significantly enriches the gravitational dynamics while maintaining the simplifying property of the reduced field equations on static, constant-curvature backgrounds. We derive the radially conserved quantity of the one-dimensional effective system, which generalizes the lower-order expressions and now involves all couplings up to quintic order in a nontrivial way. 
		
		We analyze Lifshitz backgrounds first without and then with the massive vector field. In the absence of matter, the requirement of an exact Lifshitz background imposes algebraic constraints among the cosmological constant and the quasi-topological couplings. In the matter-supported case, the vector-field amplitude \(q\) and the
		Proca mass are determined by the dynamical exponent and the curvature
		couplings; \(q\) should not be confused with an independently conserved
		Maxwell electric charge.
		Because closed-form black-hole solutions are not generally available
		for \(z\neq1\), we construct the solutions numerically. The near-horizon expansion provides regular initial data for outward
		numerical integration of the field equations, and the resulting profiles
		are checked against the required asymptotically Lifshitz behavior. As shown by the linearized analysis below, the branches
		considered here contain no physical growing mode that must be removed
		by tuning a horizon parameter.
		
		We perform the numerical analysis for \(z=1\) and \(z=2\), covering
		the three horizon topologies \(k=-1,0,+1\). We examine the normalized
		metric profile \(F(r)=f(r)/f_\infty\) for both values of \(z\), compare
		\(F(r)\) with \(g(r)\) for the \(z=2\) case, and study the full set of
		normalized profiles \(F(r)\), \(g(r)\), and
		\(H(r)=h(r)/\sqrt{f_\infty}\) for asymptotically Lifshitz black holes
		with \(z=2\).
		We also analyze the thermodynamics in detail. In addition to the
		Wald entropy and Hawking temperature, we use dimensionless logarithmic
		temperature--entropy plots to diagnose local thermal stability.
		For a fixed-theory planar \(z=2\) family, the conserved radial charge
		is used to determine the energy density, the differential and
		integrated first laws are verified numerically, and the Helmholtz
		free-energy density is evaluated relative to the Lifshitz vacuum
		normalization adopted here. In addition, we perform a controlled planar comparison
		through the genuine quartic point \(\hat\mu_5=0\), keeping the
		lower-order curvature couplings and the horizon data fixed while
		consistently recomputing the quantities constrained by the exact
		Lifshitz-background relations. This comparison shows explicitly that
		the very small quintic couplings used in the representative numerical
		examples produce profiles that are nearly indistinguishable from the
		quartic solution, whereas larger admissible values of
		\(\hat\mu_5\) generate clearly resolved changes in both the
		normalized radial profiles and the Hawking temperature. The overall
		qualitative structure of the solutions nevertheless remains consistent
		with that found previously in cubic and quartic quasi-topological
		Lifshitz gravity.
		
		The remainder of the paper is structured as follows. In Section~\ref{sec:setup} we present the quintic quasi-topological action, the Lifshitz ansatz, and the reduced field equations. Section~\ref{sec:conserved} is devoted to the derivation of the conserved quantity associated with radial evolution.
		In Section~\ref{sec:lifshitzsolutions} we discuss Lifshitz solutions both with and without matter. Section~\ref{sec:expansions} builds the near-horizon and asymptotic expansions used in the numerical analysis, and Section~\ref{sec:numerical} presents the numerical black-hole solutions for \(z=1\) and \(z=2\). In Section~\ref{sec:thermodynamics} we analyze the Wald entropy and Hawking
		temperature, determine the planar energy density from the conserved radial
		charge, verify the differential and integrated first laws for a fixed-theory
		\(z=2\) family, evaluate the corresponding Helmholtz free energy, and
		discuss both local thermal stability and the thermodynamic ensemble. Finally, Section~\ref{sec:conclusion} gives our concluding remarks.
		

		\section{Quintic quasi-topological gravity and Lifshitz ansatz}
		\label{sec:setup}
		
		\subsection{Brief review of quasi-topological gravity }
		
		We study quasi-topological gravity including curvature terms up to fifth order, coupled to a massive Abelian vector field. The action is given by
		\begin{equation}
			I = \int d^{n+1}x \sqrt{-g}\Bigl[
			-2\Lambda + \mathcal{L}_1 + \mu_2\mathcal{L}_2 + \mu_3\mathcal{L}_3
			+ \mu_4\mathcal{L}_4 + \mu_5\mathcal{L}_5
			- \frac{1}{4}F_{\mu\nu}F^{\mu\nu}
			- \frac{1}{2}m^2 A_\mu A^\mu
			\Bigr],
			\label{action}
		\end{equation}
		where \(F_{\mu\nu} = \partial_\mu A_\nu - \partial_\nu A_\mu\).
		The overall factor \(1/(16\pi G_{n+1})\) does not affect the classical
		field equations and is therefore suppressed in the derivation of the
		reduced equations below. We restore \(G_{n+1}\) explicitly when
		discussing conserved energies and thermodynamic quantities. Numerical
		thermodynamic values quoted in this work use \(G_5=1\) unless stated
		otherwise.
		The first two gravitational densities read
		\begin{equation}
			\mathcal{L}_1 = R, \qquad
			\mathcal{L}_2 = R_{abcd}R^{abcd} - 4R_{ab}R^{ab} + R^2 ,
		\end{equation}
		and \(\mathcal{L}_3\), \(\mathcal{L}_4\), \(\mathcal{L}_5\) are the cubic, quartic, and quintic quasi-topological densities, respectively.
		The cubic density is given by
		\begin{align}
			\mathcal{L}_3
			={}&
			R_a{}^c{}_b{}^d R_c{}^e{}_d{}^f R_e{}^a{}_f{}^b
			+\frac{1}{(2n-1)(n-3)}
			\bigg[
			\frac{3(3n-5)}{8}R_{abcd}R^{abcd}R
			\nonumber\\
			&-3(n-1)R_{abcd}R^{abc}{}_{e}R^{de}
			+3(n+1)R_{abcd}R^{ac}R^{bd}
			+6(n-1)R_a{}^bR_b{}^cR_c{}^a
			\nonumber\\
			&-\frac{3(3n-1)}{2}R_a{}^bR_b{}^aR
			+\frac{3(n+1)}{8}R^3
			\bigg].
			\label{L3}
		\end{align}
		The explicit forms of the quartic and quintic densities are rather lengthy; here it suffices to note that they can be expressed as linear combinations of independent curvature invariants,
		\begin{equation}
			\mathcal{L}_4 = \sum_{i=1}^{14} b_i\,\mathcal{R}^{(4)}_i,
			\qquad
			\mathcal{L}_5 = \sum_{i=1}^{24} c_i\,\mathcal{R}^{(5)}_i ,
			\label{L4L5schematic}
		\end{equation}
		with the coefficients \(b_i\) and \(c_i\) chosen so that the field equations simplify for the static constant-curvature ansatz we consider below.
		For the quartic terms we adopt the coefficients of quartic quasi-topological gravity; the quintic density is taken from the standard representative construction of quintic quasi-topological gravity \cite{DehghaniVahidinia2013,CisternaGuajardoHassaineOliva2017}.
		At quintic order, the requirement that the field equations simplify leaves one free overall choice in the coefficients. Here we adopt the standard representative quintic quasi-topological density given in Ref.~\cite{CisternaGuajardoHassaineOliva2017}; the basis of curvature invariants and our conventions are collected in Appendix~\ref{app:curvature-densities}.
		
		The dimensionless couplings \(\hat\mu_i\) appearing below are the normalized combinations of the bare couplings \(\mu_i\) that emerge upon substituting the Lifshitz ansatz into the action. The normalization is fixed by requiring that the reduced action become the polynomial in \(\Psi\) shown in Eq.~\eqref{evalaction}; throughout the analytic formulas we keep the hats to distinguish these combinations from the bare coefficients in Eq.~\eqref{action}.
		
		\subsection{Lifshitz ansatz and reduced action}
		
		We adopt the static Lifshitz ansatz
		\begin{equation}
			ds^2
			=
			-\frac{r^{2z}}{L^{2z}}\, f(r)\,dt^2
			+\frac{L^2\,dr^2}{r^2\,g(r)}
			+r^2\, d\Omega_{k,n-1}^2 ,
			\label{metlif}
		\end{equation}
		in which \(z\) is the Lifshitz dynamical exponent. 
		The metric \(d\Omega_{k,n-1}^2\) denotes an \((n-1)\)-dimensional space of constant curvature \((n-1)(n-2)k\), where \(k=-1,0,+1\) correspond, respectively, to hyperbolic, planar, and spherical horizon topologies.
		For \(k\neq0\), the constant-curvature metric can be written explicitly as
		\begin{equation}
			d\Omega_{k,n-1}^2
			=
			d\theta_1^2
			+k^{-1}\sin^2\!\bigl(\sqrt{k}\,\theta_1\bigr)
			\Bigl[
			d\theta_2^2
			+\sum_{i=3}^{n-1}
			\Bigl(\prod_{j=2}^{i-1}\sin^2\theta_j\Bigr)d\theta_i^2
			\Bigr],
		\end{equation}
		and the planar case corresponds to the appropriate \(k\to0\) limit.
		The boundary conditions for an asymptotically Lifshitz spacetime are
		\begin{equation}
			\lim_{r\to\infty} f(r) = 1, \qquad
			\lim_{r\to\infty} g(r) = 1 .
		\end{equation}
		
		We take the massive vector field to be compatible with the symmetries of the metric.
		Accordingly, we take the vector field to be of the form
		\begin{equation}
			A = A_t(r)\,dt, \qquad
			A_t(r) = q\,\frac{r^z}{L^z}\,h(r),
			\label{gfield}
		\end{equation}
		where \(q\) is a constant and, for the matter-supported Lifshitz branch, \(h(r) \to 1\) at infinity.
		
		In what follows we work exclusively in a five-dimensional bulk, setting \(n=4\), which is the natural dimension for quasi-topological constructions whose dual boundary theory is four-dimensional.
		After inserting the ansätze \eqref{metlif} and \eqref{gfield} into the action and dropping the constant transverse volume factor, the reduced action can be written in the compact form
		\begin{eqnarray}\label{evalaction}
			I_{\rm red}
			&\propto&
			\int dr\, r^{z-1}\sqrt{\frac{f}{g}}
			\Bigg[
			\left\{
			3r^4
			\left(
			-\frac{\Lambda L^2}{6}
			-\Psi
			+\hat\mu_2\Psi^2
			+\hat\mu_3\Psi^3
			+\hat\mu_4\Psi^4
			+\hat\mu_5\Psi^5
			\right)
			\right\}^{\prime} \nonumber \\
			&+& \frac{q^2r^3}{2f} \Big(g(rh'+zh)^2+m^2L^2h^2\Big)\Bigg],
		\end{eqnarray}
		where
		\begin{equation}
			\Psi(r)=g(r)-\frac{kL^2}{r^2},
			\label{PsiDef}
		\end{equation}
		and a prime indicates differentiation with respect to \(r\).
		
		Independent variation of the reduced action with respect to \(g(r)\), \(f(r)\), and \(h(r)\) produces three ordinary differential equations.
		The first field equation reads
		\begin{align}
			0={}&
			\Lambda L^2 r^{10}
			+3\hat\mu_5(3-5z)g^5 r^{10}
			+6\hat\mu_4(1-2z)g^4 r^{10}
			+(3z+3)r^{10}g
			-6z\hat\mu_2 r^{10}g^2
			\nonumber\\
			&+15\hat\mu_5L^2k(4z-3)g^4r^8
			+12\hat\mu_4L^2k(3z-2)g^3r^8
			+6z\hat\mu_2r^8L^2kg
			-3r^8L^2k
			\nonumber\\
			&-(9z-3)\hat\mu_3r^{10}g^3
			+(18z-9)\hat\mu_3r^8L^2kg^2
			+90\hat\mu_5L^4k^2(1-z)g^3r^6
			\nonumber\\
			&+36\hat\mu_4L^4k^2(1-z)g^2r^6
			-(9z-9)\hat\mu_3L^4k^2r^6g
			+30\hat\mu_5L^6k(2z-3)g^2r^4
			\nonumber\\
			&+12\hat\mu_4L^6k(z-2)gr^4
			-3\hat\mu_3L^6k^3r^4
			+6\hat\mu_4L^8k^2r^2
			+15(3-z)\hat\mu_5L^8k^2gr^2
			-9\hat\mu_5L^{10}k
			\nonumber\\
			&+g(\ln f)'
			\bigg[
			\frac{3}{2}r^{11}
			-3\hat\mu_2r^{11}g
			+3\hat\mu_2r^9L^2k
			-\frac{15}{2}\hat\mu_5r^{11}g^4
			-\frac{9}{2}\hat\mu_3r^{11}g^2
			\nonumber\\
			&\hspace{2.4cm}
			+30\hat\mu_5r^9g^3L^2k
			+9\hat\mu_3r^9gL^2k
			-\frac{9}{2}\hat\mu_3r^7L^4k^2
			-6\hat\mu_4g^3r^{11}
			\nonumber\\
			&\hspace{2.4cm}
			+18\hat\mu_4L^2kg^2r^9
			-45\hat\mu_5L^4k^2g^2r^7
			-18\hat\mu_4L^4k^2gr^7
			+6\hat\mu_4r^5L^6k
			\nonumber\\
			&\hspace{2.4cm}
			+30\hat\mu_5r^5L^6kg
			-\frac{15}{2}\hat\mu_5L^8k^2r^3
			\bigg]
			+\frac{q^2r^{10}}{4f}
			\left[
			g(rh'+zh)^2-m^2L^2h^2
			\right].
			\label{eom1}
		\end{align}
		The second equation reduces to a simple total derivative,
		\begin{equation}
			\left[
			3r^4
			\left(
			-\frac{\Lambda L^2}{6}
			-\Psi
			+\hat\mu_2\Psi^2
			+\hat\mu_3\Psi^3
			+\hat\mu_4\Psi^4
			+\hat\mu_5\Psi^5
			\right)
			\right]'
			=
			\frac{q^2r^3}{2f}
			\left[
			g(rh'+zh)^2+m^2L^2h^2
			\right].
			\label{eom2}
		\end{equation}
		Lastly, variation of the vector-field profile produces
		\begin{equation}
			2r^2h''
			-r\bigl[(\ln f)'-(\ln g)'\bigr](rh'+zh)
			+2(z+4)rh'
			+6zh
			=
			2m^2L^2\frac{h}{g}.
			\label{eom3}
		\end{equation}
		The radial evolution of the functions \(f(r)\), \(g(r)\), and \(h(r)\) is governed by Eqs.~\eqref{eom1}--\eqref{eom3}.
		In the sections that follow, we first use these equations to determine Lifshitz backgrounds and then to construct black-hole solutions that satisfy the near-horizon and asymptotic boundary conditions.

		\section{The conserved quantity}
		\label{sec:conserved}
		
		The reduced field equations admit a first integral that is conserved along the radial direction. This conserved quantity is valuable both as a check on the numerical solutions and as a concise means of connecting the near-horizon data with the asymptotic behavior of the fields.
		Even though the black-hole solutions we construct later are in five dimensions, the conserved charge admits a simple expression in \((n+1)\) dimensions.
		
		Following the approach employed in the lower-order quasi-topological Lifshitz cases, we define
		\begin{align}
			F(r) &= \frac{1}{2}\ln f(r)+z\ln\frac{r}{L}, \nonumber\\
			G(r) &= -\frac{1}{2}\ln g(r)-\ln\frac{r}{L}, \nonumber\\
			R(r) &= \ln\frac{r}{L}, \nonumber\\
			H(r) &= \ln h(r)+z\ln\frac{r}{L}.
			\label{FGHRdef}
		\end{align}
		With these definitions the metric can be expressed as
		\begin{equation}
			ds^2 = -e^{2F(r)}dt^2 + e^{2G(r)}dr^2 + L^2 e^{2R(r)} d\Omega_{k,n-1}^2 ,
			\label{FGHRmetric}
		\end{equation}
		where \(d\Omega_{k,n-1}^2\) denotes the metric of the constant-curvature \((n-1)\)-dimensional hypersurface.
		
		Upon inserting the metric \eqref{FGHRmetric} into the action and integrating by parts, the system reduces to a one-dimensional Lagrangian,
		\begin{equation}
			\mathcal{L}_{1D} = \mathcal{L}_{1g} + \mathcal{L}_{1m}.
		\end{equation}
		The gravitational part of the Lagrangian reads
		\begin{align}
			\mathcal{L}_{1g}
			={}&
			(n-1)
			\bigg[
			-\frac{2\Lambda}{n-1}e^{2G}
			+\bigl(2F'R'+(n-1)R'^2\bigr)
			\nonumber\\
			&-\frac{\hat\mu_2L^2}{3}
			\bigl(4F'R'^3+(n-4)R'^4\bigr)e^{-2G}
			\nonumber\\
			&-\frac{\hat\mu_3L^4}{5}
			\bigl(6F'R'^5+(n-6)R'^6\bigr)e^{-4G}
			\nonumber\\
			&-\frac{\hat\mu_4L^6}{7}
			\bigl(8F'R'^7+(n-8)R'^8\bigr)e^{-6G}
			\nonumber\\
			&-\frac{\hat\mu_5L^8}{9}
			\bigl(10F'R'^9+(n-10)R'^{10}\bigr)e^{-8G}
			\bigg]
			e^{F-G+(n-1)R},
			\label{L1g}
		\end{align}
		and the matter contribution is
		\begin{equation}
			\mathcal{L}_{1m}
			=
			\frac{1}{2}q^2
			\bigl(m^2+H'^2e^{-2G}\bigr)
			e^{-F+G+(n-1)R+2H}.
			\label{L1m}
		\end{equation}
		Throughout, a prime denotes differentiation with respect to \(r\).
		
		The equations derived from \(\mathcal{L}_{1D}\) can be arranged into total derivatives.
		In particular, one finds
		\begin{align}
			\mathcal{L}_{1g}-\mathcal{L}_{1m}
			={}&
			\bigg\{
			2(n-1)e^{F-G+(n-2)R}
			\bigg[
			R'
			-\frac{2}{3}\hat\mu_2L^2e^{-2G}R'^3
			-\frac{3}{5}\hat\mu_3L^4e^{-4G}R'^5
			\nonumber\\
			&\hspace{4.2cm}
			-\frac{4}{7}\hat\mu_4L^6e^{-6G}R'^7
			-\frac{5}{9}\hat\mu_5L^8e^{-8G}R'^9
			\bigg]
			\bigg\}',
			\label{sublag}
		\end{align}
		and
		\begin{align}
			\mathcal{L}_{1g}+\mathcal{L}_{1m}
			={}&
			\bigg\{
			e^{F-G+(n-1)R}
			\bigg[
			2F'+2(n-1)R'
			\nonumber\\
			&-\frac{\hat\mu_2L^2}{3}e^{-2G}
			\bigl(12F'R'^2+4(n-4)R'^3\bigr)
			\nonumber\\
			&-\frac{\hat\mu_3L^4}{5}e^{-4G}
			\bigl(30F'R'^4+6(n-6)R'^5\bigr)
			\nonumber\\
			&-\frac{\hat\mu_4L^6}{7}e^{-6G}
			\bigl(56F'R'^6+8(n-8)R'^7\bigr)
			\nonumber\\
			&-\frac{\hat\mu_5L^8}{9}e^{-8G}
			\bigl(90F'R'^8+10(n-10)R'^9\bigr)
			\bigg]
			\bigg\}'.
			\label{sumlag}
		\end{align}
		From the matter sector one obtains
		\begin{equation}
			2\mathcal{L}_{1m}
			=
			\left\{
			q^2 H' e^{-F-G+(n-1)R+2H}
			\right\}'.
			\label{matlagtotal}
		\end{equation}
		Together, Eqs.~\eqref{sublag}--\eqref{matlagtotal} combine into a total derivative, yielding a radially conserved quantity.
		Explicitly,
		\begin{align}
			\mathcal{C}_0
			={}&
			2(F'-R')e^{F-G+(n-1)R}
			\Bigl(
			1
			-2\hat\mu_2L^2R'^2e^{-2G}
			-3\hat\mu_3L^4R'^4e^{-4G}
			\nonumber\\
			&\hspace{3.6cm}
			-4\hat\mu_4L^6R'^6e^{-6G}
			-5\hat\mu_5L^8R'^8e^{-8G}
			\Bigr)
			\nonumber\\
			&-q^2H' e^{-F-G+(n-1)R+2H}.
			\label{C0FGRH}
		\end{align}
		When expressed in terms of the original functions \(f(r)\), \(g(r)\), and \(h(r)\), the conserved charge takes the form
		\begin{align}
			\mathcal{C}_0
			=
			\bigg[
			&
			\Bigl(
			1
			-2\hat\mu_2g
			-3\hat\mu_3g^2
			-4\hat\mu_4g^3
			-5\hat\mu_5g^4
			\Bigr)
			\bigl(rf'+2(z-1)f\bigr)
			\nonumber\\
			&\hspace{2.0cm}
			-q^2(zh+rh')h
			\bigg]
			\frac{r^{z+n-1}}{L^{z+1}}
			\left(\frac{g}{f}\right)^{1/2}.
			\label{Constant}
		\end{align}
		This result extends the conserved quantities known from cubic and quartic quasi-topological Lifshitz gravity. It also offers a practical consistency check for the numerical solutions that follow: once the equations of motion are satisfied, \(\mathcal{C}_0\) cannot depend on \(r\).
		
		In the AdS case \(z=1\), with the vector field turned off and \(f(r)=g(r)\), Eq.~\eqref{Constant} simplifies to
		\begin{equation}
			\mathcal{C}_0
			=
			\frac{r^{n+1}}{L^2}
			\Bigl(
			f
			-\hat\mu_2f^2
			-\hat\mu_3f^3
			-\hat\mu_4f^4
			-\hat\mu_5f^5
			\Bigr)' .
			\label{C0z1}
		\end{equation}
		This represents the quintic extension of the first integral known from
		lower-order quasi-topological black holes. For the planar \(z=1\)
		asymptotically AdS branch, the relation to the mass parameter can be
		made explicit. Defining
		\begin{equation}
			\mathcal{Q}(f)
			=
			f
			-\hat\mu_2 f^2
			-\hat\mu_3 f^3
			-\hat\mu_4 f^4
			-\hat\mu_5 f^5 ,
		\end{equation}
		Eq.~\eqref{C0z1} integrates to
		\begin{equation}
			\mathcal{Q}\bigl(f(r)\bigr)
			=
			\mathcal{Q}(1)
			-
			\frac{\mathfrak{m}}{r^n},
		\end{equation}
		where \(\mathfrak{m}\) is the AdS mass integration constant. The
		radially conserved quantity is therefore
		\begin{equation}
			\mathcal{C}_0
			=
			\frac{n\,\mathfrak{m}}{L^2}.
			\label{C0massrelation}
		\end{equation}
		Restoring the gravitational constant, the corresponding energy density
		of the planar AdS black hole is
		\begin{equation}
			\mathcal{E}
			=
			\frac{(n-1)\mathfrak{m}}
			{16\pi G_{n+1}L^2}
			=
			\frac{n-1}
			{16\pi G_{n+1}n}\,
			\mathcal{C}_0 .
			\label{C0energyrelation}
		\end{equation}
		Thus \(\mathcal{C}_0\) directly encodes the energy of the \(z=1\)
		branch. In five dimensions, Eq.~\eqref{C0energyrelation} becomes
		\begin{equation}
			\mathcal{E}
			=
			\frac{3}{64\pi G_5}\,
			\mathcal{C}_0 .
		\end{equation}
		For numerical thermodynamic quantities we set \(G_5=1\).
		
		As a numerical illustration, consider the planar \(z=1\) solution
		used below, for which
		\[
		r_0=2.13,\qquad
		(\hat\mu_2,\hat\mu_3,\hat\mu_4,\hat\mu_5)
		=
		(0.04,-0.001,0.0001,-0.0001),
		\qquad L=1.
		\]
		For this solution,
		\[
		\mathcal{Q}(1)=0.961,
		\qquad
		\mathfrak{m}=19.78070660721,
		\]
		and therefore
		\[
		\mathcal{C}_0
		=
		4\mathfrak{m}
		=
		79.12282642884.
		\]
		Restoring the gravitational normalization, the corresponding energy
		density is
		\[
		\mathcal{E}
		=
		\frac{3}{64\pi G_5}\mathcal{C}_0 .
		\]
		For the numerical convention \(G_5=1\) used for thermodynamic
		quantities in this work, this gives
		\[
		\mathcal{E}
		=
		1.18057396289549.
		\]
		The radial first integral reproduces the same value of
		\(\mathcal{C}_0\), providing a direct connection between the conserved
		quantity and the mass/energy parameter of the planar AdS solution.

		\section{Lifshitz solutions}
		\label{sec:lifshitzsolutions}
		
		In this section we work out the conditions that allow Lifshitz
		backgrounds in the quintic quasi-topological theory. As in the
		reduced field equations derived in Sec.~\ref{sec:setup}, we work
		throughout this section in five bulk dimensions and therefore set
		\(n=4\). We treat the matter-free case first and then turn on the
		massive vector field.
		
		\subsection{Matter-free solutions}
		\label{subsec:matterfree}
		
		We start by turning off the vector field,
		\begin{equation}
			h(r) = 0 .
		\end{equation}
		A planar Lifshitz background in five dimensions is given by
		\begin{equation}
			ds^2
			=
			-\frac{r^{2z}}{L^{2z}}\,dt^2
			+\frac{L^2 dr^2}{r^2}
			+r^2\sum_{i=1}^{3} d\theta_i^2 ,
			\label{matterfree-lifshitz-metric}
		\end{equation}
		which amounts to setting \(k=0\) and \(f(r)=g(r)=1\).
		Plugging this metric into the field equations yields the algebraic conditions
		\begin{equation}
			\Lambda
			=
			-\frac{2}{L^2}
			\bigl(
			2\hat\mu_5+\hat\mu_4+2-\hat\mu_2
			\bigr),
			\label{LambdaMatterFree}
		\end{equation}
		and
		\begin{equation}
			\hat\mu_3
			=
			-\frac{1}{3}
			\bigl(
			5\hat\mu_5+4\hat\mu_4-1+2\hat\mu_2
			\bigr).
			\label{mu3MatterFree}
		\end{equation}
		For \(z\neq1\), once \(\hat\mu_2\),
		\(\hat\mu_4\), and \(\hat\mu_5\) are chosen,
		Eqs.~\eqref{LambdaMatterFree} and
		\eqref{mu3MatterFree} determine the cosmological
		constant and the cubic coupling required for
		a matter-free Lifshitz vacuum of the form
		\eqref{matterfree-lifshitz-metric}.
		The case \(z=1\) is different: the geometry
		reduces to AdS, and the field equations are
		satisfied for arbitrary values of the
		quasi-topological couplings, provided that
		the cosmological constant is chosen as $
		\Lambda=-\frac{6}{L^2}
		\left(
		1-\hat\mu_2-\hat\mu_3-\hat\mu_4-\hat\mu_5
		\right) $.
		In particular, the additional coupling
		constraint \eqref{mu3MatterFree} is not
		required for \(z=1\).
		
		With Eqs.~\eqref{LambdaMatterFree} and \eqref{mu3MatterFree}, the equation following from \eqref{eom2} can be integrated once, leading to
		\begin{align}
			&
			2-\hat\mu_2+\hat\mu_4+2\hat\mu_5
			-3\Psi
			+3\hat\mu_2\Psi^2
			+\bigl(
			1-2\hat\mu_2-4\hat\mu_4-5\hat\mu_5
			\bigr)\Psi^3
			\nonumber\\
			&\hspace{3.0cm}
			+3\hat\mu_4\Psi^4
			+3\hat\mu_5\Psi^5
			=
			\frac{C}{r^4},
			\label{PsiPolynomialMatterFree}
		\end{align}
		where \(C\) is an integration constant and
		\begin{equation}
			\Psi(r)=g(r)-\frac{kL^2}{r^2}.
		\end{equation}
		When \(C=0\), the root
		\begin{equation}
			\Psi = 1
		\end{equation}
		satisfies Eq.~\eqref{PsiPolynomialMatterFree}.
		Hence,
		\begin{equation}
			g(r) = 1 + \frac{kL^2}{r^2}.
			\label{gMatterFree}
		\end{equation}
		In the hyperbolic case \(k=-1\), setting \(f(r)=g(r)\) yields
		\begin{equation}
			f(r)=g(r)=1-\frac{L^2}{r^2},
		\end{equation}
		and the metric takes the form
		\begin{equation}
			ds^2
			=
			-\frac{r^{2z}}{L^{2z}}
			\left(1-\frac{L^2}{r^2}\right)dt^2
			+
			\frac{L^2 dr^2}
			{r^2\left(1-\frac{L^2}{r^2}\right)}
			+r^2 d\Omega_{-1,3}^2 .
			\label{exactHyperbolicBH}
		\end{equation}
		This geometry possesses a horizon at \(r=L\) and constitutes the quintic generalization of the exact hyperbolic branch found in the lower-order quasi-topological Lifshitz theories.
		
		For \(C\neq0\), Eq.~\eqref{PsiPolynomialMatterFree} determines \(g(r)\) implicitly via a quintic algebraic equation.
		Unlike the cubic and quartic cases, a closed-form expression for the physical branch is not expected in general.
		Moreover, the mere existence of a real root of the polynomial does not guarantee the correct asymptotic behavior of the full metric; the remaining field equation must also be obeyed by \(f(r)\).
		Accordingly, the nontrivial black-hole solutions we are interested in will be built numerically later on.
		
		\subsection{Solutions supported by a massive vector field}
		\label{subsec:mattersolutions}
		
		Now we include the massive vector field \eqref{gfield}. For the exact Lifshitz background,
		\begin{equation}
			f(r)=g(r)=h(r)=1,
		\end{equation}
		the field equations fix algebraic relations between the vector-field amplitude \(q\), the Proca mass, the cosmological constant, and the quasi-topological couplings.
		One finds
		\begin{equation}
			q^2
			=
			\frac{
				2(z-1)
				\bigl(
				1
				-2\hat\mu_2
				-3\hat\mu_3
				-4\hat\mu_4
				-5\hat\mu_5
				\bigr)
			}{z},
			\label{q2Matter}
		\end{equation}
				\begin{equation}
			m^2
			=
			\frac{3z}{L^2},
			\label{m2Matter}
		\end{equation}
		and
		\begin{align}
			\Lambda
			=
			-\frac{1}{2L^2}
			\bigg[
			&
			\bigl(
			1
			-2\hat\mu_2
			-3\hat\mu_3
			-4\hat\mu_4
			-5\hat\mu_5
			\bigr)
			\bigl(
			z^2+2z+9
			\bigr)
			\nonumber\\
			&+
			12
			\bigl(
			\hat\mu_2
			+2\hat\mu_3
			+3\hat\mu_4
			+4\hat\mu_5
			\bigr)
			\bigg].
			\label{LambdaMatter}
		\end{align}
		
		The requirement that the vector-field amplitude \(q\) be real and nonzero imposes \(q^2>0\).
		For \(z>1\), this implies
		\begin{equation}
			1
			-2\hat\mu_2
			-3\hat\mu_3
			-4\hat\mu_4
			-5\hat\mu_5
			> 0,
		\end{equation}
		or equivalently
		\begin{equation}
			\hat\mu_2
			<
			\frac{1}{2}
			\bigl(
			1
			-3\hat\mu_3
			-4\hat\mu_4
			-5\hat\mu_5
			\bigr).
			\label{qRealityBound}
		\end{equation}
		For \(z=1\), Eq.~\eqref{q2Matter} yields \(q=0\), and the Lifshitz background reduces to the asymptotically AdS branch.
		
		With the further restriction to the conventional branch
		\(\hat\mu_2\geq0\), the allowed range of \(\Lambda\) dictated by
		\eqref{qRealityBound} becomes, in five dimensions,
		\begin{align}
			&
			-\frac{1}{2L^2}
			\bigg[
			\bigl(
			1
			-3\hat\mu_3
			-4\hat\mu_4
			-5\hat\mu_5
			\bigr)
			\bigl(
			z^2+2z+9
			\bigr)
			\nonumber\\
			&\hspace{3.0cm}
			+12
			\bigl(
			2\hat\mu_3
			+3\hat\mu_4
			+4\hat\mu_5
			\bigr)
			\bigg]
			\leq
			\Lambda
			\nonumber\\
			&\hspace{3.0cm}
			\leq
			-\frac{3}{L^2}
			\bigl(
			1+\hat\mu_3+2\hat\mu_4+3\hat\mu_5
			\bigr).
			\label{LambdaRangeMatter}
		\end{align}
		This range does not constitute an extra assumption; it merely follows
		from Eq.~\eqref{LambdaMatter} when the condition \(q^2>0\) is imposed
		on the standard branch.
		
		For \(z>1\), the algebraic relations given above ensure the existence of Lifshitz backgrounds supported by the massive vector field.
		The associated black-hole geometries call for nontrivial radial profiles \(f(r)\), \(g(r)\), and \(h(r)\) that approach unity at infinity and display the proper vanishing behavior at the horizon.
		Closed-form solutions are not available for the generic quintic theory.
		Consequently, in the following section we construct them by
		integrating the field equations outward from regular near-horizon
		series data.

		\section{Asymptotically Lifshitz black holes}
		\label{sec:expansions}
		
		We now construct black-hole solutions with nontrivial radial dependence. At the horizon the metric functions must go to zero so that the geometry possesses a regular nonextremal Killing horizon, while at infinity they must recover the Lifshitz background.
		Hence, for asymptotically Lifshitz black holes we require
		\begin{equation}
			\lim_{r\to\infty} f(r) = 1,\qquad
			\lim_{r\to\infty} g(r) = 1,\qquad
			\lim_{r\to\infty} h(r) = 1,
			\label{asymptoticbc}
		\end{equation}
		along with the condition that both \(f(r)\) and \(g(r)\) vanish linearly at the horizon \(r=r_0\).
		At \(z=1\), Eq.~\eqref{q2Matter} forces the vector-field amplitude \(q\) to vanish, which means the gauge field drops out of the gravitational sector entirely. The boundary condition on \(h(r)\) therefore matters only for the matter-supported branches, i.e., when \(z>1\).
		
		Because the generic quintic theory does not admit closed-form solutions, the black holes are obtained numerically.
		The numerical construction begins with a regular near-horizon
		expansion, which supplies initial data just outside the horizon, and
		then integrates the field equations outward. The asymptotic
		linearized analysis presented below determines whether any physical
		growing mode must be eliminated. For the branches considered in this
		work no such growing mode is present. Consequently, the two free
		coefficients appearing in the raw near-horizon expansion should not
		be interpreted as two independent shooting parameters associated
		with two asymptotic constraints. Instead, \(f_1\) fixes the residual
		normalization of the time coordinate, while for \(z>1\) the
		coefficient \(h_1\) labels the regular physical family of solutions.
		
		\subsection{Near-horizon expansion}
		\label{subsec:nearhorizon}
		
		Close to the event horizon at \(r=r_0\) we take the metric functions to vanish linearly. This motivates the expansions
		\begin{align}
			f(r)
			&=
			f_1\Bigl[
			(r-r_0)+f_2(r-r_0)^2+f_3(r-r_0)^3+\cdots
			\Bigr],
			\nonumber\\
			g(r)
			&=
			g_1(r-r_0)+g_2(r-r_0)^2+g_3(r-r_0)^3+\cdots,
			\label{near-hor}
			\\
			h(r)
			&=
			f_1^{1/2}
			\Bigl[
			h_0+h_1(r-r_0)+h_2(r-r_0)^2+h_3(r-r_0)^3+\cdots
			\Bigr].
			\nonumber
		\end{align}
		Inserting these expansions into the field equations gives
		\begin{equation}\label{h0zero}
			h_0 = 0 .
		\end{equation}
		Once \(h_0\) is known, \(g_1\) follows algebraically from \(r_0\), \(h_1\), \(z\), \(k\), \(L\), and the quasi-topological couplings.
		It is convenient to express \(g_1\) in the compact form
		\begin{equation}
			g_1
			=
			\frac{z}{r_0^3}
			\frac{\mathcal{N}_g}{\mathcal{D}_g},
			\label{g1compact}
		\end{equation}
		with
		\begin{align}
			\mathcal{N}_g
			={}&
			\hat\mu_4
			\Bigl[
			(4z^2+8z)r_0^{10}
			+12L^8k^4r_0^2
			\Bigr]
			+\hat\mu_3
			\Bigl[
			(3z^2+6z+3)r_0^{10}
			-6L^6k^3r_0^4
			\Bigr]
			\nonumber\\
			&+\hat\mu_2
			\Bigl[
			(2z^2+4z+6)r_0^{10}
			\Bigr]
			+\hat\mu_5
			\Bigl[
			(5z^2+10z-3)r_0^{10}
			-18L^{10}k^5
			\Bigr]
			\nonumber\\
			&-\bigl(z^2+2z+9\bigr)r_0^{10}
			-6L^2kr_0^8 ,
			\label{Ngdef}
		\end{align}
		and
		\begin{align}
			\mathcal{D}_g
			={}&
			\hat\mu_4
			\Bigl[
			(4h_1^2z-4h_1^2)r_0^9
			-12zL^6k^3r_0^2
			\Bigr]
			+\hat\mu_3
			\Bigl[
			(3h_1^2z-3h_1^2)r_0^9
			+9zL^4k^2r_0^4
			\Bigr]
			\nonumber\\
			&+\hat\mu_2
			\Bigl[
			(2h_1^2z-2h_1^2)r_0^9
			-6zL^2kr_0^6
			\Bigr]
			+\hat\mu_5
			\Bigl[
			(5h_1^2z-5h_1^2)r_0^9
			+15zL^8k^4
			\Bigr]
			\nonumber\\
			&+\bigl(-h_1^2z+h_1^2\bigr)r_0^9
			-3zr_0^8 .
			\label{Dgdef}
		\end{align}
		All remaining coefficients in the expansions follow recursively.
		For instance, \(h_2\), \(f_2\), and \(g_2\) are completely fixed
		once \(f_1\), \(h_1\), and the theory parameters are specified.
		The first nontrivial gauge-field coefficient is given in
		Appendix~\ref{app:near-horizon-coefficients}; all higher coefficients
		are obtained recursively by inserting the near-horizon series into
		the field equations order by order. These relations provide the
		initial data at \(r=r_0+\epsilon\).
		
		The two coefficients \(f_1\) and \(h_1\) have different physical
		roles. Under a constant rescaling of the time coordinate,
		\begin{equation}
			t_{\rm new}=c\,t_{\rm old},
		\end{equation}
		the functions transform according to
		\begin{equation}
			f_{\rm new}
			=
			\frac{f_{\rm old}}{c^2},
			\qquad
			h_{\rm new}
			=
			\frac{h_{\rm old}}{c}.
		\end{equation}
		It follows from Eq.~\eqref{near-hor} that \(f_1\) changes under this
		rescaling, whereas \(h_1\) is invariant. Thus \(f_1\) is a
		time-normalization degree of freedom rather than an independent
		physical shooting parameter. For \(z>1\), \(h_1\) is a physical
		parameter labeling the regular near-horizon family. At \(z=1\),
		\(q^2=0\), so the Proca field decouples from the gravitational
		sector and \(h_1\) does not constitute a physical gravitational
		degree of freedom.
		
		\subsection{Asymptotic behavior}
		\label{subsec:asymptoticbehavior}
		
		Far from the horizon, the black-hole solutions must approach the
		Lifshitz fixed point. We therefore write
		\begin{equation}
			f(r)=1+\delta f(r),\qquad
			g(r)=1+\delta g(r),\qquad
			h(r)=1+\delta h(r),
			\label{linearpert}
		\end{equation}
		with
		\begin{equation}
			\delta f(r),\ \delta g(r),\ \delta h(r)
			\longrightarrow 0
			\qquad
			(r\rightarrow\infty).
		\end{equation}
		
		To determine the allowed falloff modes explicitly, it is convenient
		to remove the residual normalization of the time coordinate and work
		with the invariant variables
		\begin{equation}
			G(r)=g(r),\qquad
			U(r)=\frac{h(r)}{\sqrt{f(r)}},\qquad
			V(r)=\frac{r h'(r)}{\sqrt{f(r)}}.
			\label{asymptotic-invariants}
		\end{equation}
		At the Lifshitz fixed point,
		\begin{equation}
			(G,U,V)=(1,1,0).
		\end{equation}
		Introducing
		\begin{equation}
			\delta\mathbf{X}
			=
			\begin{pmatrix}
				\delta G\\
				\delta U\\
				\delta V
			\end{pmatrix},
		\end{equation}
		and using the logarithmic radial variable
		\begin{equation}
			\rho=\ln(r/L).
		\end{equation}
		Writing the exact first-order equations for the invariant variables as
		\begin{equation}
			\frac{d\mathbf{X}}{d\rho}
			=
			\boldsymbol{\mathcal{F}}(\mathbf{X}),
			\qquad
			\mathbf{X}
			=
			\begin{pmatrix}
				G\\ U\\ V
			\end{pmatrix},
		\end{equation}
		the Lifshitz solution is the fixed point
		\[
		\boldsymbol{\mathcal{F}}(1,1,0)=0.
		\]
		The linearization matrix is therefore the Jacobian of the autonomous
		system evaluated at this fixed point,
		\begin{equation}
			M_{ab}
			=
			\left.
			\frac{\partial\mathcal{F}_a}
			{\partial X_b}
			\right|_{(G,U,V)=(1,1,0)} .
			\label{asymptoticJacobianDefinition}
		\end{equation}
		Thus, for
		\(\mathbf{X}=(1,1,0)^T+\delta\mathbf{X}\), the field equations to
		linear order take the form
		\begin{equation}
			\frac{d}{d\rho}\delta\mathbf{X}
			=
			\mathbf{M}\,\delta\mathbf{X}.
			\label{linear-asymptotic-system}
		\end{equation}
		The matrix \(\mathbf{M}\) is obtained directly by rewriting the
		first-order field equations in the variables
		\((G,U,V)\) and differentiating their autonomous right-hand sides;
		no fit to the numerical large-\(r\) solution is used in determining
		the eigenvalues.
		
		An eigenmode therefore behaves as
		\begin{equation}
			\delta\mathbf{X}
			\propto e^{\lambda\rho}
			=
			\left(\frac{r}{L}\right)^{\lambda},
			\label{asymptotic-mode-ansatz}
		\end{equation}
		so that the signs of the real parts of the eigenvalues determine
		whether the corresponding perturbations decay or grow toward the
		boundary.
		
		For the \(z=2\) representative coupling set,
		\begin{equation}
			(\hat\mu_2,\hat\mu_3,\hat\mu_4,\hat\mu_5)
			=
			\left(
			\frac{1}{25},
			-\frac{1}{1000},
			\frac{1}{10000},
			-\frac{1}{10000}
			\right),
		\end{equation}
		the characteristic polynomial of the physical invariant sector is
		\begin{equation}
			(\lambda+5)
			\left(
			\lambda^2+5\lambda+\frac{5666}{3077}
			\right)
			=0.
			\label{z2-characteristic-polynomial}
		\end{equation}
		The three physical exponents are therefore
		\begin{equation}
			\lambda_1=-5,
			\qquad
			\lambda_{2,3}
			=
			\frac{1}{2}
			\left(
			-5
			\mp
			3\sqrt{\frac{6029}{3077}}
			\right),
			\label{z2-falloff-exponents}
		\end{equation}
		or numerically
		\begin{equation}
			\lambda_{\rm phys}
			=
			\left\{
			-5,\,
			-4.599665696,\,
			-0.400334304
			\right\}.
			\label{z2-falloff-numerical}
		\end{equation}
	The two nontrivial exponents satisfy
	\begin{equation}
		\lambda_2+\lambda_3
		=-(z+3)=-5
		\qquad (z=2).
	\end{equation}
	This relation is suggestive of the source--response pairing
	familiar from holographic asymptotic expansions.
	In particular, the slowly decaying mode
	\(\lambda_{\rm slow}\simeq-0.4003\) may be interpreted
	as a source for a relevant deformation of the dual
	Lifshitz theory. Under the standard source-falloff
	identification, its effective scaling dimension would be
	\(\Delta_{\rm eff}=z+3+\lambda_{\rm slow}
	\simeq4.5997<5\).
	Since the corresponding perturbation belongs to the
	coupled gravitational and massive-vector sector, a
	definitive identification of the holographic source and
	response would require a separate analysis of the
	boundary variational problem and holographic
	renormalization.
	
	For the regular black-hole solutions considered here,
	the near-horizon parameter \(h_1\) is not fixed by the
	absence of growing radial modes. Varying \(h_1\) at
	fixed \(r_0\) can, in principle, change the coefficient
	of the slowly decaying asymptotic mode. We do not impose
	the additional condition that this coefficient vanish.
	Such a condition could provide an alternative prescription
	for selecting a particular black-hole branch and would
	be an interesting subject for future investigation.
	
		All three physical modes therefore decay as \(r\rightarrow\infty\).
		The fourth eigenvalue of the unreduced
		\(\{f,g,h,rh'\}\) system is
		\begin{equation}
			\lambda_0=0.
		\end{equation}
		This zero mode is the residual constant rescaling of the time
		coordinate discussed in connection with the near-horizon expansion,
		rather than a physical marginal perturbation. Under this rescaling,
		\(f_1\) changes, whereas the physical near-horizon parameter \(h_1\)
		and the invariant combination
		\(U=h/\sqrt{f}\) remain unchanged. Consequently, this zero eigenvalue
		does not count as an independent physical asymptotic mode and does not
		introduce an additional shooting condition. The normalization freedom
		is removed after the outward integration by fixing the asymptotic
		normalization of \(f\), equivalently by using
		\(F(r)=f(r)/f_\infty\).
		
		The curvature contribution \(kL^2/r^2\) vanishes at large radius.
		Consequently, the homogeneous \(z=2\) spectrum above is the same for
		the hyperbolic, planar, and spherical horizon topologies. We also
		repeated the linearized analysis for the additional values of the
		quintic coupling employed in the controlled comparison of
		Sec.~\ref{subsec:mu5-comparison}. The resulting spectra are
		\begin{equation}
			\begin{aligned}
				\hat\mu_5=-0.10:\quad&
				\{-5,-3.742565407,-1.257434593\},\\
				\hat\mu_5=-0.05:\quad&
				\{-5,-4.134824309,-0.865175691\},\\
				\hat\mu_5=-10^{-4}:\quad&
				\{-5,-4.599665696,-0.400334304\},\\
				\hat\mu_5=0:\quad&
				\{-5,-4.600718344,-0.399281656\},\\
				\hat\mu_5=+0.01:\quad&
				\{-5,-4.709243910,-0.290756090\}.
			\end{aligned}
			\label{mu5-asymptotic-spectra}
		\end{equation}
		Thus every \(z=2\) solution presented here has
		three physical decaying modes and no physical growing mode.
		
		The analytic mode prediction can also be checked directly against the
		long-range numerical integration. Defining the invariant residual
		\begin{equation}
			{\cal R}(r)
			=
			\left[
			(G-1)^2+(U-1)^2+V^2
			\right]^{1/2},
			\label{asymptotic-residual}
		\end{equation}
		the representative
		\(\hat\mu_5=-10^{-4}\) solution continues to decrease out to
		\(r=10^6\). A log--log fit over
		\(10^3\le r\le10^6\) gives
		\begin{equation}
			\lambda_{\rm fit}=-0.399637,
		\end{equation}
		while the effective exponent over the last radial decade,
		\(10^5\le r\le10^6\), is
		\begin{equation}
			\lambda_{\rm eff}=-0.400090420.
		\end{equation}
		Both agree closely with the independently obtained slowest
		eigenvalue
		\begin{equation}
			\lambda_{\rm slow}
			=
			-0.400334304.
		\end{equation}
		This agreement demonstrates that the numerical solution follows the
		predicted asymptotic power law rather than merely approaching the
		Lifshitz values at a finite numerical cutoff.
		
		The \(z=1\) branch must be treated separately. In this case the
		Lifshitz condition gives \(q^2=0\), so the Proca sector decouples from
		the gravitational equations. Introducing
		\begin{equation}
			\Psi(r)
			=
			g(r)-\frac{kL^2}{r^2},
		\end{equation}
		the asymptotic gravitational equation linearized around
		\(\Psi=1\) gives the homogeneous physical falloff
		\begin{equation}
			\delta\Psi(r)\propto r^{-4},
			\qquad
			\lambda_{\rm grav}=-4.
			\label{z1-falloff}
		\end{equation}
		For the three \(z=1\) parameter sets used in the numerical section,
		\(\mathcal{P}'(1)\) is nonzero, so none of these solutions lies on a
		degenerate asymptotic branch. For \(k=\pm1\), the explicit
		\(kL^2/r^2\) term supplies an additional fixed
		\(r^{-2}\) topological correction; it is not an independent
		homogeneous mode and therefore does not represent an additional
		shooting condition.
		
		The resulting mode counting is therefore unambiguous. For the
		\(z=2\) branches considered here there are three physical decaying
		modes, no physical growing modes, and one removable time-normalization
		zero mode. For \(z=1\), the physical homogeneous metric mode decays
		as \(r^{-4}\), the vector sector is absent, and again there is no
		physical growing mode. Consequently, the two coefficients
		\(f_1\) and \(h_1\) appearing in the raw near-horizon expansion
		should not be interpreted as two independent parameters required to
		eliminate two growing asymptotic modes. The role of these coefficients
		and their relation to the numerical construction are clarified below.
				
		\subsection{First-order system for numerical integration}
		\label{subsec:first-order-system}
		
		For numerical work we recast the field equations as a system of first-order ordinary differential equations.
		We define
		\begin{equation}
			j(r) = h'(r),
			\label{jdef}
		\end{equation}
		so that
		\begin{equation}
			h'(r) = j(r).
			\label{eqhprime}
		\end{equation}
		From the remaining equations one then extracts explicit expressions for \(f'(r)\), \(g'(r)\), and \(j'(r)\).
		To write the system in a compact form we introduce
		\begin{equation}
			X(r) = r j(r) + z h(r),
			\label{Xdef}
		\end{equation}
		together with
		\begin{equation}
			\Psi(r) = g(r) - \frac{k L^2}{r^2}.
			\label{PsiNumerical}
		\end{equation}
		We further define
		\begin{equation}
			\mathcal{P}(\Psi)
			=
			-\frac{\Lambda L^2}{6}
			-\Psi
			+\hat\mu_2\Psi^2
			+\hat\mu_3\Psi^3
			+\hat\mu_4\Psi^4
			+\hat\mu_5\Psi^5 .
			\label{Pnumerical}
		\end{equation}
		With these definitions the first gravitational equation takes the form
		\begin{equation}
			\mathcal{A}(r,g)
			+
			g(r)\frac{f'(r)}{f(r)}\mathcal{B}(r,g)
			+
			\frac{q^2 r^{10}}{4f(r)}
			\Bigl[
			g(r)X^2(r)-m^2L^2h^2(r)
			\Bigr]
			=
			0 .
			\label{eqnum-f}
		\end{equation}
		where
		\begin{align}
			\mathcal{A}(r,g)
			={}&
			\Lambda L^2 r^{10}
			+3\hat\mu_5(3-5z)g^5 r^{10}
			+6\hat\mu_4(1-2z)g^4 r^{10}
			+(3z+3)r^{10}g
			-6z\hat\mu_2 r^{10}g^2
			\nonumber\\
			&+15\hat\mu_5 L^2 k(4z-3)g^4 r^8
			+12\hat\mu_4 L^2 k(3z-2)g^3 r^8
			+6z\hat\mu_2 r^8 L^2 k g
			-3r^8 L^2 k
			\nonumber\\
			&-(9z-3)\hat\mu_3 r^{10}g^3
			+(18z-9)\hat\mu_3 r^8 L^2 k g^2
			+90\hat\mu_5 L^4 k^2(1-z)g^3 r^6
			\nonumber\\
			&+36\hat\mu_4 L^4 k^2(1-z)g^2 r^6
			-(9z-9)\hat\mu_3 L^4 k^2 r^6 g
			+30\hat\mu_5 L^6 k(2z-3)g^2 r^4
			\nonumber\\
			&+12\hat\mu_4 L^6 k(z-2)g r^4
			-3\hat\mu_3 L^6 k^3 r^4
			+6\hat\mu_4 L^8 k^2 r^2
			\nonumber\\
			&+15(3-z)\hat\mu_5 L^8 k^2 g r^2
			-9\hat\mu_5 L^{10} k .
			\label{Adefnumerical}
		\end{align}
		and
		\begin{align}
			\mathcal{B}(r,g)
			={}&
			\frac{3}{2}r^{11}
			-3\hat\mu_2 r^{11}g
			+3\hat\mu_2 r^9 L^2 k
			-\frac{15}{2}\hat\mu_5 r^{11}g^4
			-\frac{9}{2}\hat\mu_3 r^{11}g^2
			\nonumber\\
			&+30\hat\mu_5 r^9 g^3 L^2 k
			+9\hat\mu_3 r^9 g L^2 k
			-\frac{9}{2}\hat\mu_3 r^7 L^4 k^2
			-6\hat\mu_4 g^3 r^{11}
			\nonumber\\
			&+18\hat\mu_4 L^2 k g^2 r^9
			-45\hat\mu_5 L^4 k^2 g^2 r^7
			-18\hat\mu_4 L^4 k^2 g r^7
			+6\hat\mu_4 r^5 L^6 k
			\nonumber\\
			&+30\hat\mu_5 r^5 L^6 k g
			-\frac{15}{2}\hat\mu_5 L^8 k^2 r^3 .
			\label{Bdefnumerical}
		\end{align}
		The second gravitational equation can be expressed as
		\begin{equation}
			\frac{d}{dr}
			\Bigl[
			3r^4\mathcal{P}(\Psi)
			\Bigr]
			=
			\frac{q^2 r^3}{2f(r)}
			\Bigl[
			g(r)X^2(r) + m^2 L^2 h^2(r)
			\Bigr],
			\label{eqnum-g}
		\end{equation}
		while the equation for the vector field takes the form
		\begin{equation}
			2r^2 j'(r)
			-r
			\left(
			\frac{f'(r)}{f(r)}
			-
			\frac{g'(r)}{g(r)}
			\right)
			X(r)
			+2(z+4)r j(r)
			+6z h(r)
			=
			\frac{2m^2 L^2 h(r)}{g(r)} .
			\label{eqnum-j}
		\end{equation}
		The outward numerical integration is based on the set of equations
		\eqref{eqhprime}, \eqref{eqnum-f}, \eqref{eqnum-g}, and \eqref{eqnum-j}.
		In practice, one solves them algebraically for \(h'(r)\), \(f'(r)\), \(g'(r)\), and \(j'(r)\), which gives four coupled first-order equations governing the variables
		\begin{equation}
			\{f(r),\,g(r),\,h(r),\,j(r)\}.
		\end{equation}
		
		The integration begins at \(r=r_0+\epsilon\), with \(\epsilon\)
		a small positive number, and the initial values of
		\(f\), \(g\), \(h\), and \(j\) are read from the near-horizon
		series \eqref{near-hor}. Since \(f_1\) represents the residual
		normalization of the time coordinate, one may use a convenient raw
		normalization during the integration. After the solution has been
		obtained, the time coordinate is rescaled so that
		\begin{equation}
			\lim_{r\to\infty}f(r)=1.
		\end{equation}
		For \(z>1\), a representative value of the invariant horizon
		parameter \(h_1\) selects a member of the regular solution family.
		
		The asymptotic conditions are then tested using the invariant
		quantities introduced in Eq.~\eqref{asymptotic-invariants}. In
		particular, a genuine asymptotically Lifshitz solution must satisfy
		\begin{equation}
			g(r)\longrightarrow1,
			\qquad
			\frac{h(r)}{\sqrt{f(r)}}\longrightarrow1,
		\end{equation}
		with the residual perturbation decaying according to the physical
		eigenvalues obtained from the linearized system
		\eqref{linear-asymptotic-system}. Thus the identification of an
		asymptotically Lifshitz branch does not rely solely on agreement with
		the background values at a finite numerical cutoff.

		\section{Numerical solutions}
		\label{sec:numerical}
		
		We now turn to the numerical black-hole solutions that follow from the
		first-order system introduced in
		Sec.~\ref{subsec:first-order-system}. Starting the integration just
		outside the horizon with the near-horizon expansion, we evolve the
		regular initial data outward and verify that the invariant
		perturbations approach the Lifshitz fixed point with the falloff
		exponents derived in Sec.~\ref{subsec:asymptoticbehavior}. For the
		branches displayed here, the physical asymptotic spectrum contains no
		growing mode, so no horizon parameter is tuned to cancel an unstable
		asymptotic mode. The residual normalization of the time coordinate is
		fixed after the outward integration, as described below. We treat the
		two representative cases \(z=1\) and \(z=2\) separately, and for each
		dynamical exponent examine all three horizon topologies
		\(k=-1,0,+1\). This choice is not a structural limitation of the
		numerical construction. The reduced field equations, the near-horizon
		expansion, and the outward integration procedure are formulated for
		generic \(z\), so the same method can be applied to other admissible
		values of the dynamical exponent. For each chosen \(z\), however, the
		Lifshitz-background parameters must be recomputed and the corresponding
		asymptotic linear spectrum must be examined. In particular, the Proca
		reality condition and the nondegeneracy conditions of the gravitational
		branch must be satisfied, and if physical growing asymptotic modes are
		present, the appropriate regular near-horizon parameter must be tuned
		by shooting. Thus there is no fundamental obstruction to generic \(z\),
		although some regions of parameter space can require a more involved
		numerical analysis than the representative \(z=1\) and \(z=2\) cases
		studied here. The parameter choices for the numerical runs are summarized in
		Table~\ref{tab:numerical-parameters} and indicated on the corresponding
		figure panels.
		Throughout the figures, the symbols \(\mu_i\) refer to the dimensionless couplings \(\hat\mu_i\) employed in the main text.
		
		\begin{table}[t]
			\centering
			\caption{
				Parameter values and near-horizon data adopted for the numerical
				solutions shown in the figures. In the figure panels, \(\mu_i\)
				denotes the dimensionless coupling \(\hat\mu_i\) introduced in the
				text. The column \(f_1\) gives the coefficient of \(r-r_0\) in the
				near-horizon expansion of the asymptotically normalized metric
				function, for which \(f(\infty)=1\). For the \(z=2\) integrations
				the raw time-coordinate normalization was chosen as
				\(f_1^{\rm raw}=1\); the values of \(f_1\) listed below are obtained
				after the subsequent normalization at infinity. The invariant
				near-horizon family parameter is \(h_1=1\) for all displayed
				\(z=2\) solutions. At \(z=1\), \(q^2=0\) and the Proca field
				decouples, so \(h_1\) is not applicable.
			}
			\label{tab:numerical-parameters}
			\resizebox{\textwidth}{!}{%
				\begin{tabular}{c c c c c c c c c}
					\hline
					Figure & \(z\) & \(k\) & \(r_0\) & \(\hat\mu_2\) &
					\(\hat\mu_3\) & \(\hat\mu_4\) & \(\hat\mu_5\) &
					\(\{f_1,h_1\}\) \\
					\hline
					Fig.~\ref{fig:f-z1}(a) & 1 & \(-1\) & 1.68 &
					0.4 & \(-0.1\) & \(-0.001\) & 0.001 &
					\(\{1.657380916,\text{--}\}\) \\
					
					Fig.~\ref{fig:f-z1}(b) & 1 & \(0\) & 2.13 &
					0.04 & \(-0.001\) & 0.0001 & \(-0.0001\) &
					\(\{1.804694836,\text{--}\}\) \\
					
					Fig.~\ref{fig:f-z1}(c) & 1 & \(+1\) & 2.91 &
					0.4 & \(-0.01\) & 0.001 & \(-0.001\) &
					\(\{0.8399302251,\text{--}\}\) \\
					\hline
					
					Fig.~\ref{fig:f-z2}(a) & 2 & \(-1\) & 1.5 &
					0.04 & \(-0.001\) & 0.0001 & \(-0.0001\) &
					\(\{101.1967545,1\}\) \\
					
					Fig.~\ref{fig:f-z2}(b) & 2 & \(0\) & 3.0 &
					0.04 & \(-0.001\) & 0.0001 & \(-0.0001\) &
					\(\{2.054518192,1\}\) \\
					
					Fig.~\ref{fig:f-z2}(c) & 2 & \(+1\) & 4.0 &
					0.04 & \(-0.001\) & 0.0001 & \(-0.0001\) &
					\(\{0.1952797050,1\}\) \\
					\hline
					
					Fig.~\ref{fig:fg-z2}(a) & 2 & \(-1\) & 1.5 &
					0.04 & \(-0.001\) & 0.0001 & \(-0.0001\) &
					\(\{101.1967545,1\}\) \\
					
					Fig.~\ref{fig:fg-z2}(b) & 2 & \(0\) & 2.0 &
					0.04 & \(-0.001\) & 0.0001 & \(-0.0001\) &
					\(\{32.94898500,1\}\) \\
					
					Fig.~\ref{fig:fg-z2}(c) & 2 & \(+1\) & 4.0 &
					0.04 & \(-0.001\) & 0.0001 & \(-0.0001\) &
					\(\{0.1952797050,1\}\) \\
					\hline
					
					Fig.~\ref{fig:fgh-z2}(a) & 2 & \(-1\) & 1.5 &
					0.04 & \(-0.001\) & 0.0001 & \(-0.0001\) &
					\(\{101.1967545,1\}\) \\
					
					Fig.~\ref{fig:fgh-z2}(b) & 2 & \(0\) & 3.0 &
					0.04 & \(-0.001\) & 0.0001 & \(-0.0001\) &
					\(\{2.054518192,1\}\) \\
					
					Fig.~\ref{fig:fgh-z2}(c) & 2 & \(+1\) & 4.0 &
					0.04 & \(-0.001\) & 0.0001 & \(-0.0001\) &
					\(\{0.1952797050,1\}\) \\
					\hline
				\end{tabular}%
			}
		\end{table}
		
\begin{figure}[t]
			\centering
			\begin{minipage}{0.32\textwidth}
				\centering
				\includegraphics[width=\linewidth,height=0.23\textheight,keepaspectratio]{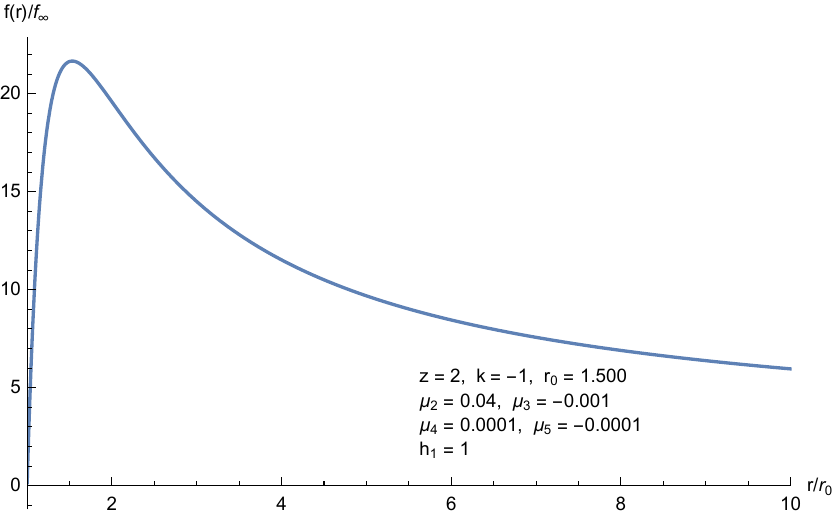}
				\par\vspace{1mm}
				{\small (a) \(k=-1\)}
			\end{minipage}
			\hfill
			\begin{minipage}{0.32\textwidth}
				\centering
				\includegraphics[width=\linewidth,height=0.23\textheight,keepaspectratio]{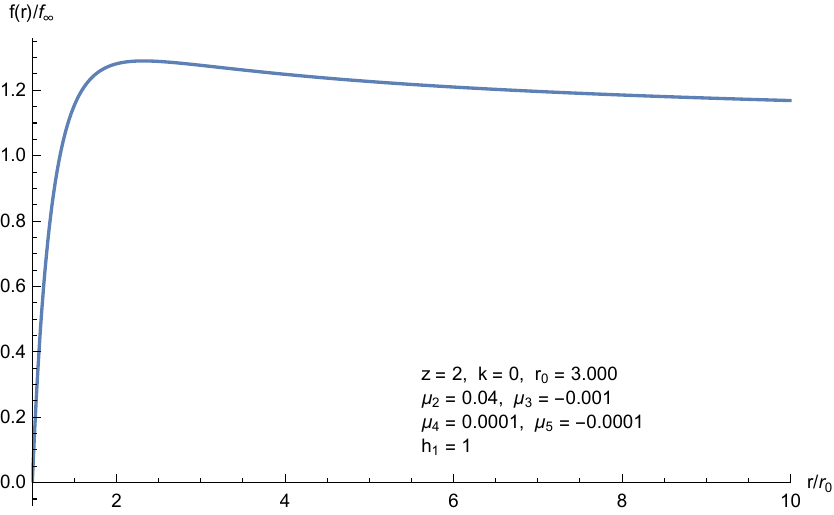}
				\par\vspace{1mm}
				{\small (b) \(k=0\)}
			\end{minipage}
			\hfill
			\begin{minipage}{0.32\textwidth}
				\centering
				\includegraphics[width=\linewidth,height=0.23\textheight,keepaspectratio]{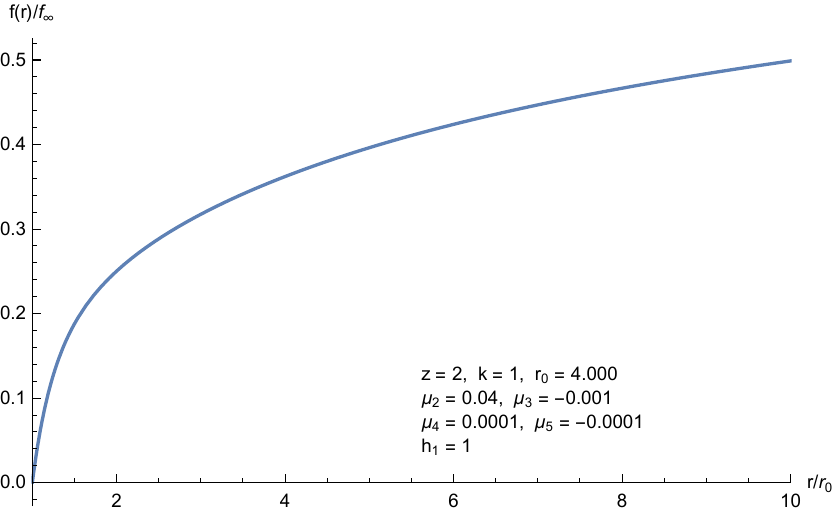}
				\par\vspace{1mm}
				{\small (c) \(k=+1\)}
			\end{minipage}
			\caption{
				Normalized metric function
				\(F(r)=f(r)/f_\infty\) for the \(z=2\) Lifshitz branch, shown for
				the three horizon topologies \(k=-1,0,+1\) as a function of
				\(r/r_0\). The coupling values and the choice \(h_1=1\) are
				indicated on the respective panels. In every case \(F(r)\)
				vanishes at the horizon and approaches unity asymptotically.
				The hyperbolic solution develops a pronounced intermediate
				overshoot, the planar solution exhibits a much milder one,
				whereas the spherical example approaches its asymptotic value
				from below without a comparable intermediate maximum.
			}
			\label{fig:f-z2}
		\end{figure}

\begin{figure}[t]
			\centering
			\begin{minipage}{0.32\textwidth}
				\centering
				\includegraphics[width=\linewidth,height=0.23\textheight,keepaspectratio]{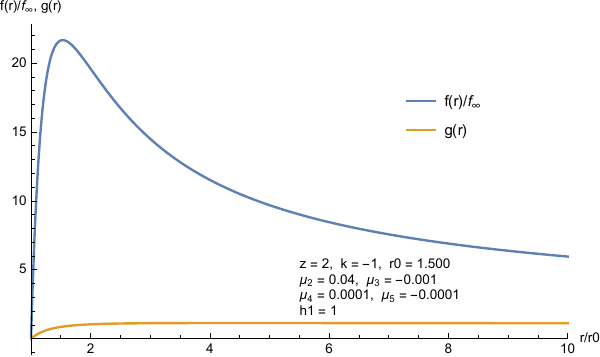}
				\par\vspace{1mm}
				{\small (a) \(k=-1\)}
			\end{minipage}
			\hfill
			\begin{minipage}{0.32\textwidth}
				\centering
				\includegraphics[width=\linewidth,height=0.23\textheight,keepaspectratio]{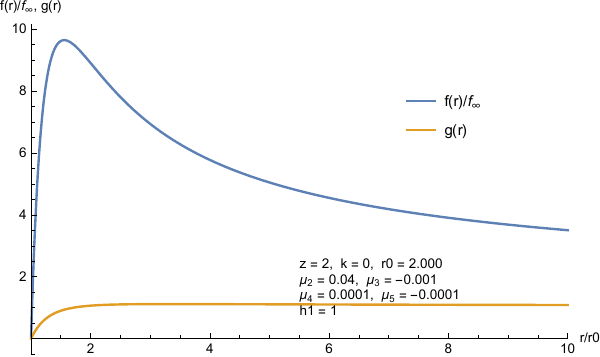}
				\par\vspace{1mm}
				{\small (b) \(k=0\)}
			\end{minipage}
			\hfill
			\begin{minipage}{0.32\textwidth}
				\centering
				\includegraphics[width=\linewidth,height=0.23\textheight,keepaspectratio]{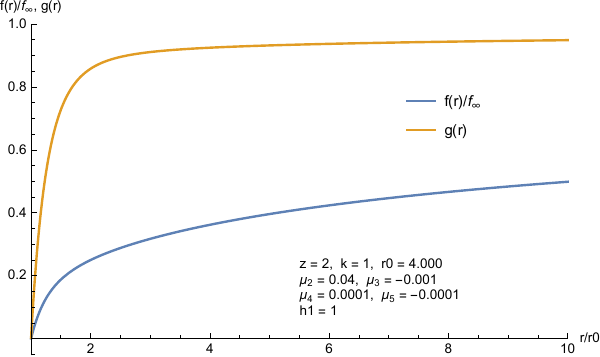}
				\par\vspace{1mm}
				{\small (c) \(k=+1\)}
			\end{minipage}
			\caption{
				Comparison of the normalized metric function
				\(F(r)=f(r)/f_\infty\) and \(g(r)\) for the \(z=2\) Lifshitz
				branch, shown for \(k=-1,0,+1\) as functions of \(r/r_0\).
				The parameter choices and the fixed value \(h_1=1\) are indicated
				on the individual panels. Both functions vanish at the horizon
				and approach unity asymptotically, but they need not coincide at
				intermediate radii. The hyperbolic case exhibits the strongest
				enhancement of \(F(r)\), while the corresponding variation of
				\(g(r)\) is substantially milder.
			}
			\label{fig:fg-z2}
		\end{figure}

\begin{figure}[t]
			\centering
			\begin{minipage}{0.32\textwidth}
				\centering
				\includegraphics[width=\linewidth,height=0.23\textheight,keepaspectratio]{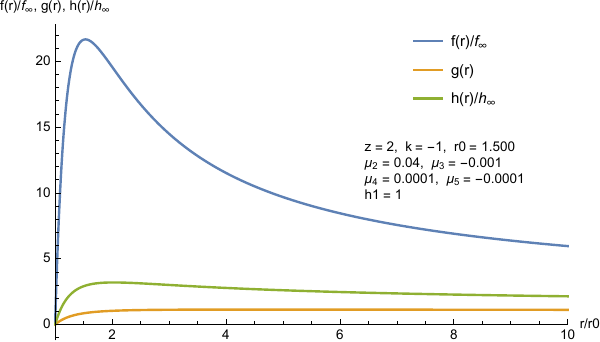}
				\par\vspace{1mm}
				{\small (a) \(k=-1\)}
			\end{minipage}
			\hfill
			\begin{minipage}{0.32\textwidth}
				\centering
				\includegraphics[width=\linewidth,height=0.23\textheight,keepaspectratio]{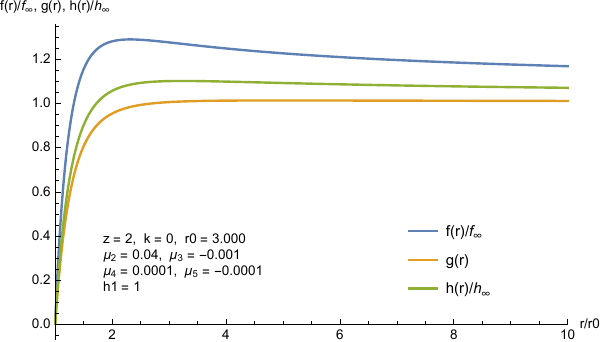}
				\par\vspace{1mm}
				{\small (b) \(k=0\)}
			\end{minipage}
			\hfill
			\begin{minipage}{0.32\textwidth}
				\centering
				\includegraphics[width=\linewidth,height=0.23\textheight,keepaspectratio]{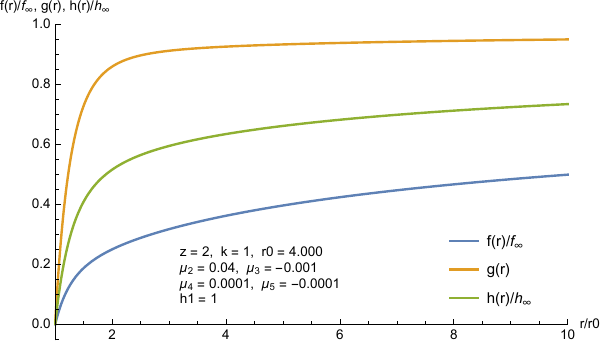}
				\par\vspace{1mm}
				{\small (c) \(k=+1\)}
			\end{minipage}
			\caption{
				Full set of normalized profiles
				\(F(r)=f(r)/f_\infty\), \(g(r)\), and
				\(H(r)=h(r)/h_\infty\), with
				\(h_\infty=\sqrt{f_\infty}\), for the \(z=2\) Lifshitz branch.
				The three horizon topologies \(k=-1,0,+1\) are shown as functions
				of the dimensionless radius \(r/r_0\). The parameter values and
				the fixed choice \(h_1=1\) are indicated within each panel.
				All three profiles vanish at the horizon and approach unity
				asymptotically. The hyperbolic solution exhibits the largest
				intermediate deformation, the planar case shows a milder one,
				and the spherical example approaches the Lifshitz values from
				below without a comparable intermediate maximum.
			}
			\label{fig:fgh-z2}
		\end{figure}

		We set \(L=1\) in all numerical examples. Given a choice of \(z\)
		and the quasi-topological couplings, the Lifshitz conditions
		\eqref{q2Matter}--\eqref{LambdaMatter} determine \(q\), \(m\), and
		\(\Lambda\). These quantities are therefore constrained by the
		asymptotic Lifshitz background rather than being independent inputs
		of the numerical construction. For the matter-supported \(z=2\)
		profile integrations displayed in Figs.~\ref{fig:f-z2}--
		\ref{fig:fgh-z2}, the second-order near-horizon expansion is used
		to initialize the integration at
		\[
		r_{\mathrm{start}}=r_0+\epsilon,
		\qquad
		\epsilon=10^{-6},
		\]
		and the solutions are evolved outward to
		\(r_{\max}=10^6\). The calculations are performed with
		50-digit arithmetic. The adaptive integration is required to meet
		an absolute accuracy target of approximately \(10^{-30}\) and a
		relative precision target of approximately \(10^{-25}\), with no
		preset restriction on the number of integration steps. An adaptive
		algorithm capable of treating stiff regions of the differential
		system is employed when necessary. 		
		These numerical choices were used for the \(z=2\)
		profiles and provide stable large-radius behavior over the range
		relevant to the asymptotic analysis.
		
		The asymptotic normalization constant $f_\infty$ is
		extracted from the raw numerical solution by
		extrapolating the large-radius behavior
		$f(r)=f_\infty+A(r/L)^{\lambda_{\rm slow}}+\cdots$,
		using the slowest physical exponent obtained in
		Sec.~\ref{subsec:asymptoticbehavior}. For the
		representative $z=2$, $k=0$, $r_0=3$ solution,
		comparison of extrapolations using different
		large-radius pairs up to $r=10^6$ gives an estimated
		relative uncertainty of order $4\times10^{-5}$
		in $f_\infty$.
		
		The \(z=1\) examples belong to the metric-only \(q^2=0\) branch,
		so the Proca horizon datum \(h_1\) plays no role in their
		numerical construction. For these solutions the
		first-order metric equation can be integrated directly from the
		horizon, so no finite horizon offset \(\epsilon\) is required. The
		integration is performed over
		\[
		r_0\leq r\leq 100\,r_0
		\]
		using 40-digit arithmetic, with absolute and relative numerical
		accuracy targets of approximately \(10^{-25}\) and \(10^{-20}\),
		respectively, and with no preset restriction on the number of
		integration steps.
		
		The coefficients \(f_1\) and \(h_1\) appearing in the near-horizon
		expansion have different roles. The coefficient \(f_1\) fixes the
		residual normalization of the time coordinate: under a constant
		rescaling of time it changes, and after the outward integration this
		freedom is removed by imposing \(f(\infty)=1\). By contrast, for the
		matter-supported \(z>1\) branches, \(h_1\) is invariant under this
		normalization and labels the regular physical family of near-horizon
		solutions. At \(z=1\), the Lifshitz condition gives \(q^2=0\), so the
		Proca sector decouples from the gravitational equations and \(h_1\)
		does not constitute a physical gravitational degree of freedom.
		
		The linearized analysis of
		Sec.~\ref{subsec:asymptoticbehavior} shows that all of the branches
		displayed here contain no physical growing asymptotic mode.
		Consequently, neither \(f_1\) nor \(h_1\) is adjusted in order to
		cancel an unstable mode. For the controlled \(z=2\), \(k=0\)
		comparison introduced below, we choose the convenient raw
		normalization \(f_1^{\rm raw}=1\) and hold the physical family parameter
		\(h_1=1\) fixed while varying \(\hat\mu_5\). The residual time
		normalization is removed after integration by imposing
		\(f(\infty)=1\).
		The parameter sets in Table~\ref{tab:numerical-parameters} were chosen
		as representative regular black-hole solutions and were not intended
		as a controlled scan of the fifth-order coupling. In particular, the
		small values of \(|\hat\mu_5|\) used there make the quintic solutions
		very close to the quartic limit. To isolate the fifth-order
		contribution quantitatively, we therefore perform a dedicated comparison in which \(\hat\mu_5\) alone is
		varied among the independent gravitational couplings.
		
		The aim of the numerical examples shown here is to establish that
		quintic quasi-topological Lifshitz black holes exist, and to illustrate
		their qualitative behavior for a representative set of coupling
		values. A comprehensive exploration of the full quintic coupling space
		lies outside the scope of this paper.
		
		As an independent check of the numerical integration, we also evaluate
		the radially conserved quantity \(\mathcal{C}_0\) of
		Eq.~\eqref{Constant} along a representative \(z=2\), \(k=0\)
		solution. The parameters are
		\[
		L=1,\qquad r_0=3,\qquad h_1=1,
		\]
		and
		\[
		(\hat\mu_2,\hat\mu_3,\hat\mu_4,\hat\mu_5)
		=
		(0.04,-0.001,0.0001,-0.0001).
		\]
		For this consistency test the same solution was reintegrated with
		80-digit working precision, and \(\mathcal{C}_0\) was evaluated
		directly at points of the numerical integration mesh. Representative
		values are shown in Table~\ref{tab:C0-conservation}.
		
		\begin{table}[t]
			\centering
			\caption{
				Numerical conservation of the raw conserved charge
				\(\mathcal{C}_0^{\rm(raw)}\) for the representative
				\(z=2\), \(k=0\) Lifshitz black hole.
				The relative deviation is measured with respect to the
				near-horizon mesh value. The maximum relative drift over
				the displayed radial interval is approximately
				\(5.3\times10^{-34}\).
			}
			\label{tab:C0-conservation}
			\begin{tabular}{c c c}
				\hline
				\(r\) &
				\(\mathcal{C}_0^{\rm(raw)}(r)\) &
				\(
				|\Delta\mathcal{C}_0^{\rm(raw)}|
				/
				|\mathcal{C}_0^{\rm(raw)}|
				\)
				\\
				\hline
				3.0100986   & 807.703717164504 & \(0\) \\
				4.9981328   & 807.703717164504 & \(1.1\times10^{-40}\) \\
				9.8595660   & 807.703717164504 & \(1.0\times10^{-40}\) \\
				29.4943483  & 807.703717164504 & \(1.2\times10^{-38}\) \\
				98.4886096  & 807.703717164504 & \(4.0\times10^{-36}\) \\
				202.5779107 & 807.703717164504 & \(1.1\times10^{-34}\) \\
				300.0000000 & 807.703717164504 & \(5.3\times10^{-34}\) \\
				\hline
			\end{tabular}
		\end{table}
		It is important to distinguish the raw integration normalization from the
		boundary-normalized one used for physical thermodynamic quantities.  The
		values displayed in Table~\ref{tab:C0-conservation} are evaluated directly
		from the raw numerical solution, for which the residual time normalization
		has not yet been removed.  If
		\begin{equation}
			F(r)=\frac{f(r)}{f_\infty},
			\qquad
			H(r)=\frac{h(r)}{\sqrt{f_\infty}},
		\end{equation}
		then the conserved charge transforms according to
		\begin{equation}
			\mathcal{C}_0^{\rm(norm)}
			=
			\frac{\mathcal{C}_0^{\rm(raw)}}{\sqrt{f_\infty}} .
			\label{C0Normalization}
		\end{equation}
		Accordingly, the value quoted in Table~\ref{tab:C0-conservation} is the raw
		conserved charge, whereas the thermodynamic analysis below uses
		\(\mathcal{C}_0^{\rm(norm)}\), associated with the asymptotic normalization
		\(F(\infty)=1\).  The two values therefore refer to the same numerical
		solution expressed in different time normalizations.
		
		The conserved charge therefore remains constant throughout the
		integration to substantially higher accuracy than required for the
		radial profiles. This provides a direct numerical consistency check
		of the first-order integration and of the near-horizon initial data.
		
		We begin with the \(z=1\) branch. Here the Lifshitz scaling collapses
		to the relativistic AdS case, and the massive-vector amplitude \(q\)
		vanishes. The \(z=1\) branch thus offers a useful consistency check on the
		numerical construction before one proceeds to the genuinely Lifshitz
		geometries with \(z\neq1\).
		
\begin{figure}[t]
			\centering
			\begin{minipage}{0.32\textwidth}
				\centering
				\includegraphics[width=\linewidth,height=0.23\textheight,keepaspectratio]{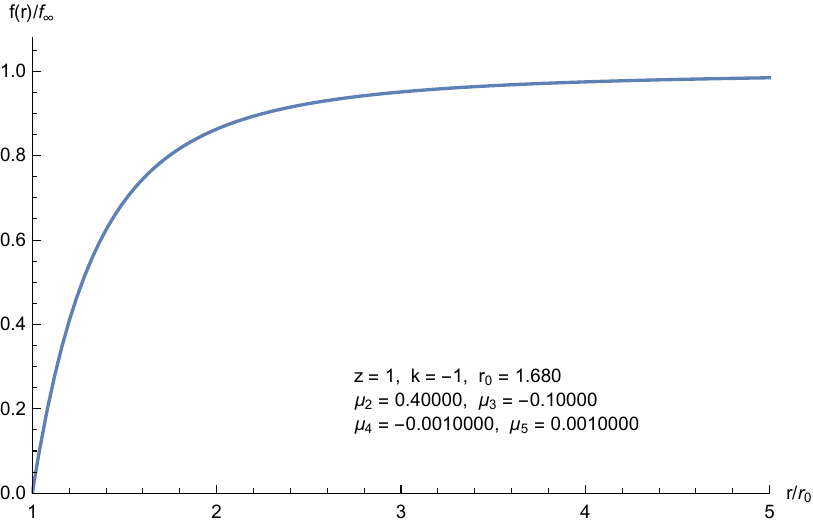}
				\par\vspace{1mm}
				{\small (a) \(k=-1\)}
			\end{minipage}
			\hfill
			\begin{minipage}{0.32\textwidth}
				\centering
				\includegraphics[width=\linewidth,height=0.23\textheight,keepaspectratio]{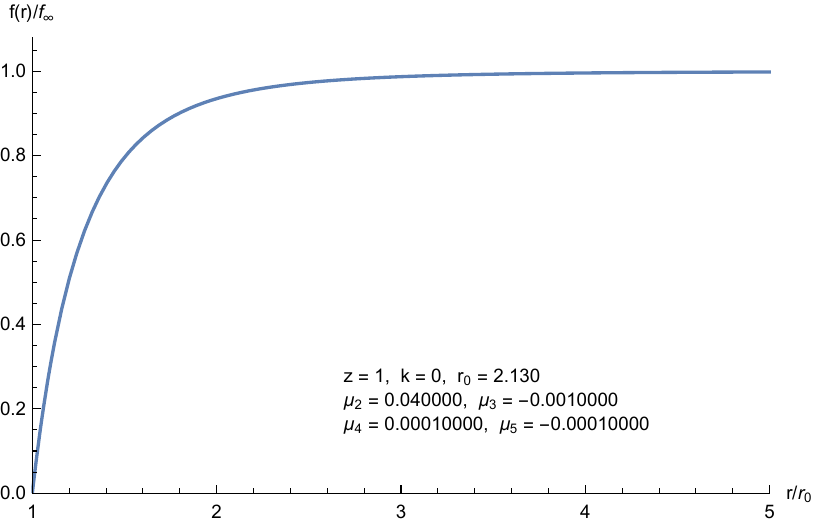}
				\par\vspace{1mm}
				{\small (b) \(k=0\)}
			\end{minipage}
			\hfill
			\begin{minipage}{0.32\textwidth}
				\centering
				\includegraphics[width=\linewidth,height=0.23\textheight,keepaspectratio]{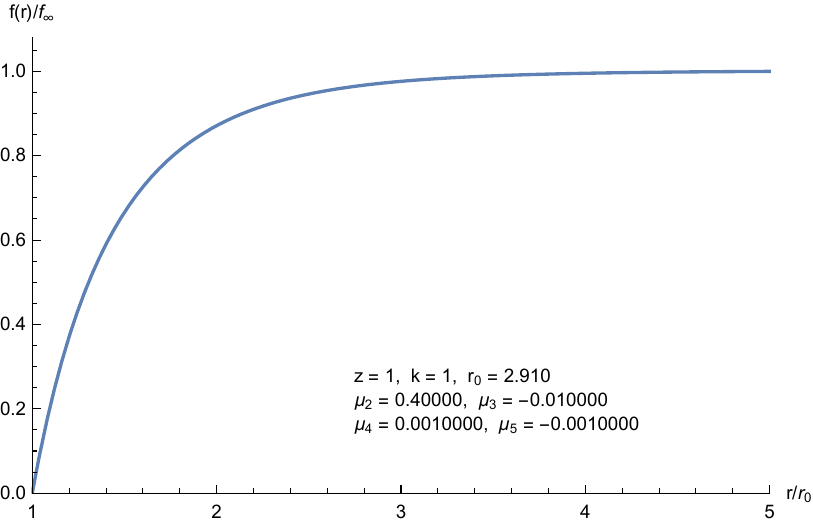}
				\par\vspace{1mm}
				{\small (c) \(k=+1\)}
			\end{minipage}
			\caption{
				Normalized metric function
				\(F(r)=f(r)/f_\infty\) for the \(z=1\) branch, shown for the
				three horizon topologies \(k=-1,0,+1\) as a function of the
				dimensionless radius \(r/r_0\). The coupling parameters adopted
				in each panel are given on the plots. For these asymptotically
				AdS solutions \(f_\infty=1\), so the normalized profile is
				numerically identical to \(f(r)\). In every case \(F(r)\)
				vanishes at the horizon and approaches unity at large \(r\).
			}
			\label{fig:f-z1}
		\end{figure}

		Figure~\ref{fig:f-z1} shows the normalized metric function
		\[
		F(r)=\frac{f(r)}{f_\infty}
		\]
		for the three horizon topologies, plotted against the dimensionless
		radius \(r/r_0\). For the \(z=1\) solutions considered here the
		asymptotically AdS condition gives \(f_\infty=1\), so \(F(r)\) is
		numerically identical to the corresponding \(f(r)\). In every case
		shown, \(F(r)\) vanishes at the horizon and approaches unity at large
		radius.
		
		The detailed shape of the radial profile varies with the horizon
		curvature and the choice of quasi-topological couplings, but the
		overall behavior matches what is seen in lower-order
		quasi-topological black holes.
		
		We now move to the genuine Lifshitz case \(z=2\).
		Figure~\ref{fig:f-z2} displays the normalized metric profile
		\[
		F(r)=\frac{f(r)}{f_\infty}
		\]
		for the three topologies \(k=-1,0,+1\), plotted against the
		dimensionless radius \(r/r_0\). For the regular-family solutions shown
		here we fix \(h_1=1\), while the residual time-coordinate normalization
		is removed after integration through \(f_\infty\).
		
		Relative to the \(z=1\) solutions, the \(z=2\) profiles exhibit a more
		pronounced dependence on the horizon topology. In the hyperbolic case,
		\(F(r)\) develops a strong intermediate overshoot before relaxing
		toward its asymptotic value, while the planar solution shows a much
		milder overshoot. By contrast, for the spherical example displayed
		here, \(F(r)\) rises toward its asymptotic value from below and does
		not develop a comparable intermediate maximum. Thus the detailed
		radial behavior depends sensitively on both the horizon topology and
		the chosen horizon radius.		
		
		Figure~\ref{fig:fg-z2} provides a closer look at the metric sector for
		\(z=2\), displaying the normalized function
		\(F(r)=f(r)/f_\infty\) together with \(g(r)\) as functions of
		\(r/r_0\). Both functions vanish at the horizon and approach unity
		asymptotically, but they are not required to coincide at intermediate
		radii. This distinction is intrinsic to the Lifshitz ansatz, since the
		reduced field equations do not generically impose \(f(r)=g(r)\).
		
		The profiles also show that the magnitude of the radial
		variation can differ substantially between the two metric functions.
		For the hyperbolic example, \(F(r)\) develops a pronounced
		intermediate overshoot whereas the corresponding variation of
		\(g(r)\) is considerably milder. The planar case exhibits a more
		moderate departure from the asymptotic values, while in the spherical
		example both profiles rise toward the Lifshitz boundary values without
		a comparable large overshoot. Thus the two metric functions respond
		differently to the same topology and coupling choice even though they
		share the same horizon and asymptotic boundary conditions.
		
		For the planar example in
		Fig.~\ref{fig:fg-z2}(b), we choose
		$r_0=2$ to illustrate a more pronounced
		separation between the normalized metric
		functions $F(r)$ and $g(r)$.
		By contrast, the planar examples in
		Figs.~\ref{fig:f-z2}(b) and
		\ref{fig:fgh-z2}(b) use $r_0=3$.
		These correspond to distinct regular
		black-hole solutions with the same
		gravitational couplings and $h_1=1$;
		their radial profiles are therefore not
		intended to represent the same solution
		across the three figures.
		
		Figure~\ref{fig:fgh-z2} presents the full numerical solution for the
		\(z=2\) branch in terms of the normalized profiles
		\[
		F(r)=\frac{f(r)}{f_\infty},
		\qquad
		g(r),
		\qquad
		H(r)=\frac{h(r)}{h_\infty},
		\qquad
		h_\infty=\sqrt{f_\infty},
		\]
		plotted as functions of \(r/r_0\). Since the Proca field is essential
		for sustaining the Lifshitz asymptotics when \(z\neq1\), the profile
		\(H(r)\) forms an integral part of the black-hole solution rather than
		an independent probe.
		
		For the solutions displayed here, all three normalized functions are
		regular outside the horizon and approach unity asymptotically. Their
		intermediate radial behavior, however, depends strongly on the horizon
		topology. The hyperbolic example exhibits the largest departure from
		the Lifshitz values, with a pronounced enhancement of \(F(r)\) and a
		smaller but clearly visible response in \(H(r)\) and \(g(r)\). The
		planar solution shows substantially milder overshoots. By contrast,
		for the spherical example the three profiles rise toward their
		asymptotic values from below over the displayed radial interval and do
		not develop a comparable intermediate maximum.
		
		\subsection{Controlled dependence on the quintic coupling}
		\label{subsec:mu5-comparison}
		
		The representative solutions above establish the existence and
		qualitative structure of the quintic Lifshitz black holes, but the
		small values of \(\hat\mu_5\) used in those examples do not by
		themselves resolve the fifth-order contribution from the corresponding
		quartic theory. We therefore perform a controlled comparison through
		the genuine quartic point \(\hat\mu_5=0\).
		
		We fix
		\begin{equation}
			(n,z,k,L,r_0)=(4,2,0,1,3),
			\qquad
			(\hat\mu_2,\hat\mu_3,\hat\mu_4)
			=(0.04,-0.001,0.0001),
			\qquad
			h_1=1,
		\end{equation}
		and vary only
		\begin{equation}
			\hat\mu_5=-0.10,\;-0.05,\;-10^{-4},\;0,\;+0.01.
		\end{equation}
	    This five-point parameter scan is designed to quantify the dependence of the geometry and the thermodynamic observables on the quintic coupling while all other independent gravitational and  horizon parameters are held fixed.
		Here ``vary only \(\hat\mu_5\)'' refers to the independent
		gravitational inputs. The Proca amplitude \(q^2\) and the cosmological
		constant \(\Lambda\) are constrained by the exact Lifshitz-background
		relations and are consequently recomputed at each value of
		\(\hat\mu_5\).
		
		For this parameter set the standard real Proca-supported background
		requires \(q^2>0\), corresponding to
		\(\hat\mu_5<0.18452\). In addition, the asymptotic spectrum develops
		a growing physical mode on the positive-\(\hat\mu_5\) side above
		\(\hat\mu_5\simeq0.033896\). The values used in the comparison below
		were therefore chosen inside the same regular-family sector. For all
		five cases the outward integrations remain regular, the metric
		functions stay positive outside the horizon, the relevant
		higher-curvature denominator remains nonzero, and the solutions
		approach the Lifshitz fixed point.
		
		The raw numerical integrations use \(f_1=1\). Because the residual
		time-coordinate normalization generally gives
		\(f(r)\rightarrow f_\infty\), we compare the physically normalized
		functions
		\begin{equation}
			F(r)=\frac{f(r)}{f_\infty},
			\qquad
			g(r),
			\qquad
			H(r)=\frac{h(r)}{h_\infty},
			\qquad
			h_\infty=\sqrt{f_\infty}.
		\end{equation}
		Thus \(F\), \(g\), and \(H\) all approach unity at the Lifshitz
		boundary. Notice that no independent \(g_\infty\) normalization is
		introduced: \(g(r)\rightarrow1\) is itself part of the asymptotic
		Lifshitz boundary condition.
		
\begin{figure}[t]
			\centering
			\includegraphics[width=\textwidth]{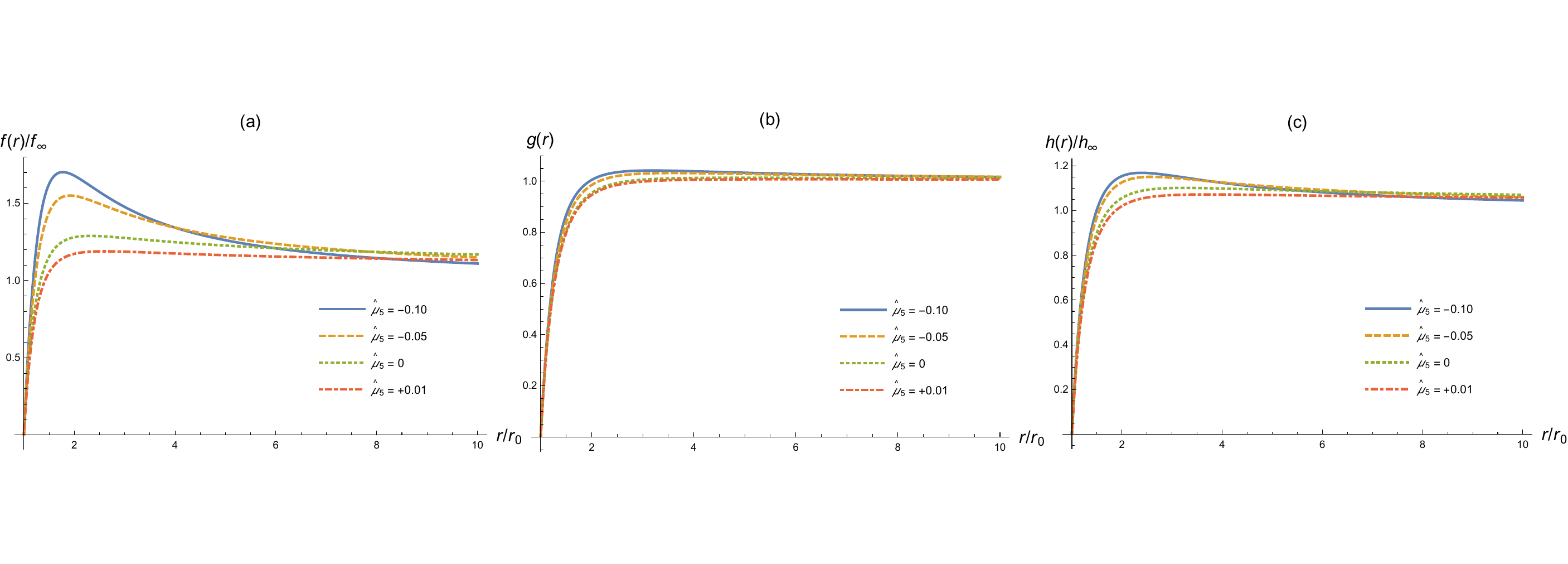}
			\caption{
				Controlled dependence of the normalized \(z=2\), \(k=0\)
				Lifshitz black-hole solution on the quintic coupling
				\(\hat\mu_5\). The fixed parameters are
				\(L=1\), \(r_0=3\),
				\((\hat\mu_2,\hat\mu_3,\hat\mu_4)
				=(0.04,-0.001,0.0001)\), and \(h_1=1\).
				The panels show
				(a) \(F(r)=f(r)/f_\infty\),
				(b) \(g(r)\), and
				(c) \(H(r)=h(r)/h_\infty\), where
				\(h_\infty=\sqrt{f_\infty}\).
				The case \(\hat\mu_5=0\) is the genuine quartic reference.
				The larger admissible values of \(|\hat\mu_5|\) produce
				clearly resolved changes in the radial profiles, while all
				displayed solutions approach the same normalized Lifshitz
				asymptotics.
			}
			\label{fig:mu5-profiles}
		\end{figure}

		Figure~\ref{fig:mu5-profiles} shows the resulting profiles.
		Changing the quintic coupling produces a clearly resolved deformation
		once \(|\hat\mu_5|\) is increased beyond the very small values used
		in the representative examples. Relative to the genuine quartic
		solution, the largest absolute change in \(F\) is approximately
		\(0.474\) for \(\hat\mu_5=-0.10\), \(0.295\) for
		\(\hat\mu_5=-0.05\), and \(0.108\) for
		\(\hat\mu_5=+0.01\). The corresponding changes in \(g\) and \(H\)
		are smaller but clearly resolved.
		
		By contrast, the representative small-coupling value
		\(\hat\mu_5=-10^{-4}\) is quantitatively very close to the quartic
		limit. We obtain
		\begin{equation}
			\max|\Delta F|=9.01\times10^{-4},
			\qquad
			\max|\Delta g|=7.71\times10^{-5},
			\qquad
			\max|\Delta H|=2.76\times10^{-4},
			\label{mu5-small-profile-difference}
		\end{equation}
		where each difference is taken with respect to the
		\(\hat\mu_5=0\) solution. Hence the near-overlap of the profiles in
		the representative examples is a consequence of the small
		quintic coupling, rather than evidence that the fifth-order
		interaction has no numerical effect.

	\begin{table}[t]
		\centering
		\caption{
			Controlled comparison with the genuine quartic solution
			\(\hat\mu_5=0\). The parameters
			\(n=4\), \(z=2\), \(k=0\), \(L=1\), \(r_0=3\),
			\((\hat\mu_2,\hat\mu_3,\hat\mu_4)
			=(0.04,-0.001,0.0001)\), and \(h_1=1\)
			are held fixed. The quantities \(q^2\) and \(\Lambda\)
			are recomputed from the exact Lifshitz-background
			conditions for each value of \(\hat\mu_5\).
			The profile differences are evaluated relative to the
			quartic solution.
		}
		\label{tab:mu5-comparison}
		\resizebox{\textwidth}{!}{%
			\begin{tabular}{c c c c c c c c}
				\hline
				\(\hat\mu_5\) &
				\(q^2\) &
				\(\Lambda L^2\) &
				\(\max|\Delta F|\) &
				\(\max|\Delta g|\) &
				\(\max|\Delta H|\) &
				\(T/r_0^2\) &
				\(\Delta T/T_0~(\%)\) \\
				\hline
				
				\(-0.10\) &
				1.4226 &
				-9.9219 &
				0.474363 &
				0.0619122 &
				0.107019 &
				0.4256366 &
				\(+12.2981\) \\
				
				\(-0.05\) &
				1.1726 &
				-8.9969 &
				0.294698 &
				0.0341932 &
				0.0733769 &
				0.4098217 &
				\(+8.1255\) \\
				
				\(-10^{-4}\) &
				0.9231 &
				-8.07375 &
				\(9.00751\times10^{-4}\) &
				\(7.70886\times10^{-5}\) &
				\(2.76337\times10^{-4}\) &
				0.3791327 &
				\(+0.02871\) \\
				
				\(0\) &
				0.9226 &
				-8.0719 &
				0 &
				0 &
				0 &
				0.3790239 &
				0 \\
				
				\(+0.01\) &
				0.8726 &
				-7.8869 &
				0.108166 &
				0.00795216 &
				0.0360571 &
				0.3651301 &
				\(-3.6657\) \\
				
				\hline
			\end{tabular}%
		}
	\end{table}

\begin{figure}[t]
	\centering
	\includegraphics[width=\textwidth]{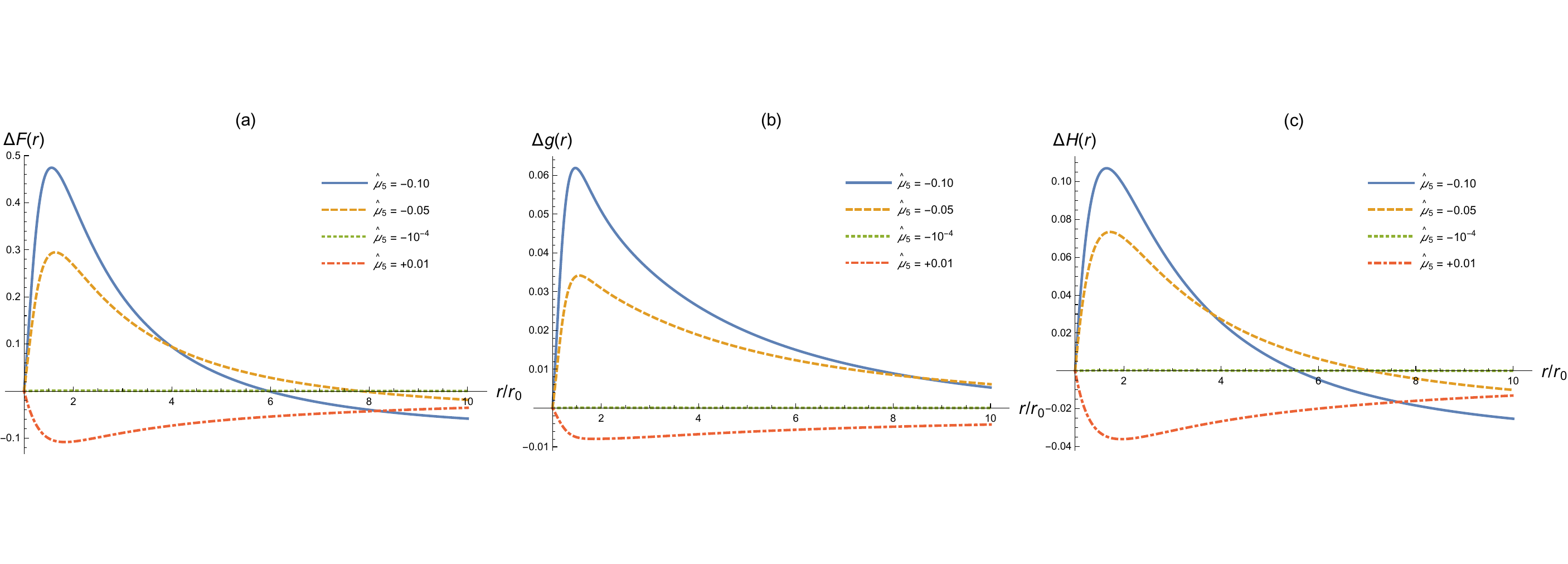}
	\caption{
		Direct differences between the quintic solutions and the
		genuine quartic reference \(\hat\mu_5=0\), for the same
		controlled \(z=2\), \(k=0\) comparison as in
		Fig.~\ref{fig:mu5-profiles}.
		The panels show
		(a) \(\Delta F(r)=F(r;\hat\mu_5)-F(r;0)\),
		(b) \(\Delta g(r)=g(r;\hat\mu_5)-g(r;0)\), and
		(c) \(\Delta H(r)=H(r;\hat\mu_5)-H(r;0)\).
		The small-coupling choice
		\(\hat\mu_5=-10^{-4}\) remains extremely close to the
		quartic solution, whereas the larger admissible values
		\(\hat\mu_5=-0.05\), \(-0.10\), and \(+0.01\)
		produce clearly resolved fifth-order corrections.
		All displayed differences decay as the corresponding
		solutions approach the common Lifshitz asymptotics.
	}
	\label{fig:mu5-differences}
\end{figure}

		These numerical results demonstrate that the quintic theory supports
		regular asymptotically Lifshitz black-hole solutions for the
		representative parameter choices explored here. The representative examples illustrate the existence and qualitative structure of the solutions but do not constitute a complete survey of the quintic coupling space. In particular, the
		small values of \(|\hat\mu_5|\) used in the representative profile plots
		naturally produce solutions that lie very close to the corresponding
		quartic theory.
		
		The controlled comparison presented above makes the fifth-order
		effect quantitative. Keeping
		\(n,z,k,L,r_0,\hat\mu_2,\hat\mu_3,\hat\mu_4\) and \(h_1\) fixed,
		while varying only \(\hat\mu_5\) among the independent gravitational
		couplings, produces clearly resolved changes in the normalized radial
		profiles for larger admissible values of the quintic coupling.
		Conversely, the case \(\hat\mu_5=-10^{-4}\) differs from the genuine
		quartic solution by less than \(10^{-3}\) in the largest normalized
		profile deviation. The near-overlap of these profiles is
		therefore a consequence of the small quintic coupling chosen in those
		illustrative examples, rather than an indication that the fifth-order
		interaction has no effect.
		
		The quintic polynomial also requires a specification of the gravitational
		branch. The solutions considered here are chosen on the branch satisfying
		\(\Psi\to1\) at infinity and are continued continuously through the genuine
		quartic point \(\hat\mu_5=0\). This is the Einstein-connected branch: when
		the higher-curvature couplings are taken to zero, the polynomial reduces
		continuously to the Einstein--Proca result. A local merger of algebraic
		branches would require
		\[
		\mathcal{P}'(\Psi)
		=
		-1+2\hat\mu_2\Psi
		+3\hat\mu_3\Psi^2
		+4\hat\mu_4\Psi^3
		+5\hat\mu_5\Psi^4
		=
		0.
		\]
		We therefore monitor \(\mathcal{P}'(\Psi(r))\) throughout the numerical
		exterior. 
		For the five values
		\begin{equation}
			\hat\mu_5=-0.10,\,-0.05,\,-10^{-4},\,0,\,+0.01,
		\end{equation}
		the minimum values of
		\(\left|\mathcal{P}'\right|\) along the corresponding solutions are
		approximately \(0.9798\), \(0.9747\), \(0.9221\), \(0.9216\), and
		\(0.8704\), respectively, and the sign remains the Einstein-connected
		one in every case. Hence none of the displayed solutions encounters a
		branch collision.
		
		The algebraic quintic can nevertheless possess additional real roots.
		For the fixed lower-order couplings used in the controlled comparison,
		two additional real sheets appear as the coupling is varied, while the
		physical asymptotic root \(\Psi=1\) itself becomes degenerate only at
		\(\hat\mu_5\simeq0.18452\). For example, at
		\(\hat\mu_5=0.01\) the equation
		\(\mathcal{P}(\Psi)=\mathcal{P}(1)\) has the three real roots
		\(\Psi\simeq-3.4577\), \(1\), and \(2.7195\), whereas the corresponding
		black-hole solution explores only
		\(1.22\times10^{-6}\lesssim\Psi(r)\lesssim1.0078\).
		Thus the existence of additional algebraic roots does not imply that
		the numerical solution changes branch. A comprehensive classification
		of all regular black-hole branches over the full quintic coupling space
		remains beyond the scope of the present work.
		
		In the next section we compute the thermodynamic quantities for these
		numerical branches and investigate their local thermal behavior.
		
		\section{Thermodynamics}
		\label{sec:thermodynamics}
		
		We now study the thermodynamic properties of the quintic quasi-topological Lifshitz black holes. The temperature follows from the surface gravity at the horizon, and the entropy is obtained via the Wald formula, which is the appropriate prescription for higher-curvature theories.
		In the thermodynamic discussion that follows, all entropy values are given as densities per unit horizon volume and are expressed in the same gravitational units used for the action.
		
		The entropy is computed using the Iyer--Wald prescription for higher-curvature gravity,
		\begin{equation}
			S
			=
			-2\pi
			\int_{\Sigma_h} d^{\,n-1}x\,\sqrt{\tilde g}\,
			Y^{abcd}\hat\epsilon_{ab}\hat\epsilon_{cd},
			\qquad
			Y^{abcd}
			=
			\frac{\partial\mathcal{L}}{\partial R_{abcd}},
			\label{WaldEntropy}
		\end{equation}
		where \(\Sigma_h\) is a spatial cross section of the horizon, \(\tilde g\) is the determinant of the induced metric on \(\Sigma_h\), and \(\hat\epsilon_{ab}\) is the horizon binormal.
		
		For the static black-hole solutions under consideration, the entropy density \(\mathcal{S}=S/V_{n-1}\) reads
		\begin{align}
			\mathcal{S}
			=
			\frac{r_0^{\,n-1}}{4G_{n+1}}
			\bigg[
			1
			&-2k\hat\mu_2
			\frac{(n-1)L^2}{(n-3)r_0^2}
			+3k^2\hat\mu_3
			\frac{(n-1)L^4}{(n-5)r_0^4}
			\nonumber\\
			&-4k\hat\mu_4
			\frac{(n-1)L^6}{(n-7)r_0^6}
			+5k^2\hat\mu_5
			\frac{(n-1)L^8}{(n-9)r_0^8}
			\bigg].
			\label{EntropyDensity}
		\end{align}
		The signs appearing in Eq.~\eqref{EntropyDensity} are a consequence of the normalization adopted for the dimensionless couplings \(\hat\mu_i\) in the reduced action \eqref{evalaction}.
		In this convention, one retrieves the lower-order results simply by taking the appropriate higher-curvature couplings to vanish.
		Even though the numerical solutions presented in this paper are built in a five-dimensional bulk (\(n=4\)), Eq.~\eqref{EntropyDensity} has been expressed in general \((n+1)\)-dimensional form to make transparent the contribution of each curvature order to the Wald entropy.
		The five-dimensional entropy density appearing in the numerical plots follows from Eq.~\eqref{EntropyDensity} by taking \(n=4\), with the same coupling conventions adopted in the reduced action.
		
		The Hawking temperature follows from the requirement that the Euclidean section be regular at the horizon. This yields
		\begin{equation}
			T
			=
			\left.
			\frac{r^{z+1}\sqrt{f'(r)g'(r)}}{4\pi L^{z+1}}
			\right|_{r=r_0}.
			\label{Temperature}
		\end{equation}
		Given a numerical black-hole solution, one obtains the temperature from Eq.~\eqref{Temperature} using \(r_0\) and the horizon values of \(f'(r_0)\) and \(g'(r_0)\), while Eq.~\eqref{EntropyDensity} supplies the entropy density.
		The controlled quintic comparison of
		Sec.~\ref{subsec:mu5-comparison} also provides a direct
		thermodynamic measure of the fifth-order effect. Since this
		comparison is planar, \(k=0\), and is performed at fixed
		\(r_0=3\), Eq.~\eqref{EntropyDensity} reduces to
		\begin{equation}
			\mathcal{S}
			=
			\frac{r_0^3}{4G_5}
			=
			\frac{27}{4G_5}.
		\end{equation}
		so that the entropy density is independent of \(\hat\mu_5\) in
		this particular fixed-\(r_0\) comparison. For the numerical values
		quoted below we set \(G_5=1\), giving \(\mathcal{S}=27/4\). The Hawking temperature,
		however, changes appreciably. Relative to the genuine quartic point
		\(\hat\mu_5=0\), the temperature shifts are
		\begin{equation}
			\frac{\Delta T}{T_0}
			=
			\begin{cases}
				+12.30\%, & \hat\mu_5=-0.10,\\
				+8.13\%,  & \hat\mu_5=-0.05,\\
				+0.0287\%, & \hat\mu_5=-10^{-4},\\
				-3.67\%,  & \hat\mu_5=+0.01.
			\end{cases}
		\end{equation}
		These values agree with the geometric comparison in
		Table~\ref{tab:mu5-comparison}: the small-coupling choice
		\(\hat\mu_5=-10^{-4}\) produces an almost imperceptible correction,
		whereas larger admissible quintic couplings lead to clearly resolved
		changes in the thermal response.
		
		\subsection{Energy normalization and thermodynamic ensemble}
		\label{subsec:energy-thermo}
		
		We now adopt the energy normalization associated with the conserved
		radial charge \(\mathcal C_0\).  We fix the residual time
		normalization by imposing \(f(\infty)=1\), and for the planar branch
		we take the exact Lifshitz vacuum to define the zero of the energy.
		Following the normalization used in the corresponding lower-order
		quasi-topological Lifshitz constructions, a homogeneous
		five-dimensional Lifshitz state satisfies
		\begin{equation}
			\frac{\mathcal C_0}{16\pi G_5}
			=
			\mathcal E+p,
			\qquad
			z\mathcal E=3p,
			\label{BoundaryChargeWard}
		\end{equation}
		where \(\mathcal E\) and \(p\) are respectively the energy density
		and pressure.  It follows that
		\begin{equation}
			\mathcal E
			=
			\frac{3}{z+3}
			\frac{\mathcal C_0}{16\pi G_5}.
			\label{EnergyFromC0}
		\end{equation}
		This normalization is consistent with the lower-order
		quasi-topological Lifshitz constructions
		\cite{BrennaDehghaniMann2011,GhanaatianBazrafshanBrenna2014}.
		
		For a planar horizon, the same conserved quantity can be evaluated
		directly at \(r=r_0\).  Using
		\[
		f(r)=f_1(r-r_0)+\cdots,
		\qquad
		g(r)=g_1(r-r_0)+\cdots,
		\]
		together with Eq.~\eqref{Temperature}, one obtains
		\begin{equation}
			\mathcal C_0
			=
			\frac{r_0^{z+4}}{L^{z+1}}
			\sqrt{f_1g_1}
			=
			16\pi G_5\,T\mathcal S,
			\qquad
			\mathcal S=\frac{r_0^3}{4G_5}
			\quad (k=0).
			\label{C0HorizonTs}
		\end{equation}
		Equations~\eqref{EnergyFromC0} and \eqref{C0HorizonTs} therefore
		yield the planar Smarr relation
		\begin{equation}
			\mathcal E
			=
			\frac{3}{z+3}\,T\mathcal S,
			\qquad
			\mathcal F
			\equiv
			\mathcal E-T\mathcal S
			=
			-\frac{z}{z+3}\,T\mathcal S
			=
			-\frac{z}{z+3}
			\frac{\mathcal C_0}{16\pi G_5},
			\label{PlanarSmarrFreeEnergy}
		\end{equation}
		where \(\mathcal F\) is the Helmholtz free-energy density.
		
		The thermodynamic family considered below keeps \(z\), \(L\), the
		curvature couplings, and the leading Lifshitz Proca source fixed.
		In this ensemble there is no independent Maxwell-like electric charge
		varied along the family and hence no separate \(\Phi\,dQ\) term in
		the first law.  The reason for this, together with the distinction
		between Maxwell and Proca radial fluxes, is discussed explicitly in
		Subsec.~\ref{subsec:first-law-test}.
		
		\begin{figure}[t]
			\centering
			\includegraphics[width=0.55\textwidth]{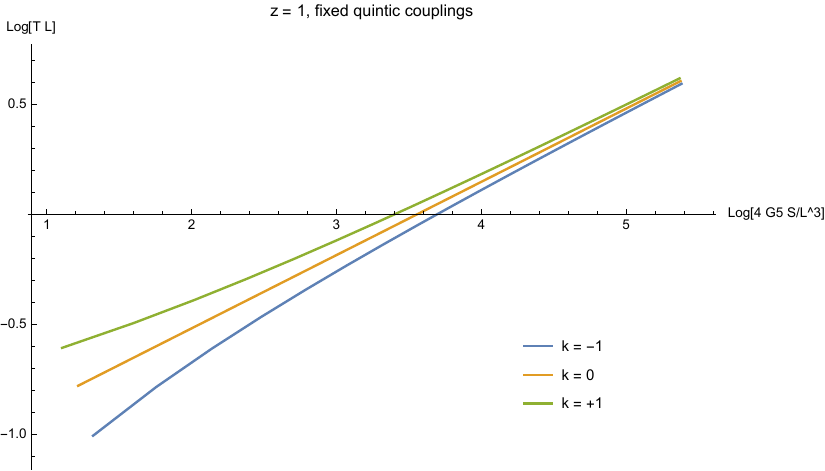}
			\caption{
				Dimensionless logarithmic temperature--entropy relation for the
				\(z=1\) branch and the three horizon topologies
				\(k=-1,0,+1\), at fixed quintic couplings
				\(\hat\mu_2=0.04\), \(\hat\mu_3=-0.001\),
				\(\hat\mu_4=10^{-4}\), and
				\(\hat\mu_5=-10^{-4}\).
				The horizontal axis labeled
				\(\mathrm{Log}[4G_5\mathcal{S}/L^3]\) and the vertical axis labeled
				\(\mathrm{Log}[TL]\) both represent natural logarithms.
				For the planar branch the numerical slope is
				\(d\ln(TL)/d\ln(4G_5\mathcal{S}/L^3)=1/3\), as expected
				analytically. All three displayed branches have positive
				slope and hence positive heat capacity over the range shown.
			}
			\label{fig:st-z1}
		\end{figure}
		
		\begin{figure}[t]
			\centering
			\includegraphics[width=0.55\textwidth]{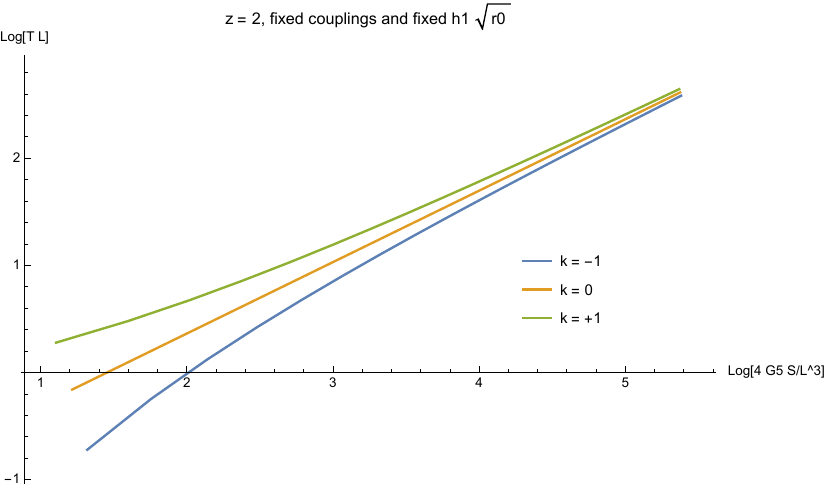}
			\caption{
				Dimensionless logarithmic temperature--entropy relation for the
				\(z=2\) Lifshitz branch and the three horizon topologies
				\(k=-1,0,+1\), at fixed quintic couplings
				\(\hat\mu_2=0.04\), \(\hat\mu_3=-0.001\),
				\(\hat\mu_4=10^{-4}\), and
				\(\hat\mu_5=-10^{-4}\).
				The family is constructed at fixed
				\(h_1\sqrt{r_0}=\sqrt{3}\), with \(r_0/L\) varied along
				each topology branch. The horizontal axis labeled
				\(\mathrm{Log}[4G_5\mathcal{S}/L^3]\) and the vertical axis labeled
				\(\mathrm{Log}[TL]\) both represent natural logarithms. The planar branch is nearly linear with
				\(d\ln(TL)/d\ln(4G_5\mathcal{S}/L^3)
				\simeq 0.666667\), in agreement with the expected value
				\(2/3\). The \(k=\pm1\) branches show the curvature induced
				by the horizon topology, while all three displayed branches
				retain positive slope and therefore positive heat capacity
				over the range shown.
			}
			\label{fig:st-z2}
		\end{figure}
		
		With the black-hole solutions in hand, the Hawking temperature is
		obtained from the surface gravity at the horizon, while the entropy
		density is evaluated using the Wald prescription appropriate to the
		higher-curvature theory. To probe the local thermal behavior, we use
		the dimensionless logarithmic variables
		\begin{equation}
			X
			=
			\ln\!\left(\frac{4G_5\mathcal{S}}{L^3}\right),
			\qquad
			Y
			=
			\ln(TL).
			\label{dimensionlessThermoLogs}
		\end{equation}
		Here and throughout the thermodynamic discussion, \(\ln\) denotes
		the natural logarithm. The ``Log'' notation appearing on the axes of
		Figs.~\ref{fig:st-z1} and \ref{fig:st-z2} likewise denotes the
		natural logarithm, as in the numerical implementation.
		The numerical values shown on these logarithmic axes should therefore
		not be interpreted as powers of ten. For example,
		\(\mathrm{Log}[TL]=3\) corresponds to \(TL=e^3\simeq20.1\).
		The different locations of the \(z=1\) and \(z=2\) families on the
		temperature axis arise primarily from their different horizon-radius
		scalings rather than from a change of temperature units or
		normalization. In particular, for the planar branches considered
		below, \(T\propto r_0\) at \(z=1\), whereas
		\(T\propto r_0^2\) at \(z=2\), while in both cases
		\(\mathcal{S}\propto r_0^3\). The resulting logarithmic slopes are
		therefore \(1/3\) and \(2/3\), respectively, as verified explicitly
		by the numerical families below.
		
		The heat-capacity density is
		\begin{equation}
			C
			=
			T\frac{d\mathcal{S}}{dT}.
		\end{equation}
		Since the constant factors in \(X\) and \(Y\) disappear upon
		differentiation, the slope of the curves displayed in
		Figs.~\ref{fig:st-z1} and \ref{fig:st-z2} is
		\begin{equation}
			\frac{dY}{dX}
			=
			\frac{d\log T}{d\log\mathcal{S}}
			=
			\frac{\mathcal{S}}{C}.
			\label{logSlopeHeatCapacity}
		\end{equation}
		Consequently, for the positive-entropy branches considered here, a
		positive slope is equivalent to a positive heat capacity and therefore
		provides a direct diagnostic of local thermal stability.
		
		Figure~\ref{fig:st-z1} shows the \(z=1\) branch.  In this limit
		\(q^2=0\), the massive vector decouples, and the solutions reduce to
		the relativistic AdS sector.  For the planar case,
		\(T\propto r_0\) and \(\mathcal{S}\propto r_0^3\), so that
		\begin{equation}
			\frac{d\log T}{d\log\mathcal{S}}
			=
			\frac{1}{3},
		\end{equation}
		in exact agreement with the numerical slope.  The curved-horizon
		branches are also monotonic over the displayed interval: the local
		logarithmic slopes range from approximately \(0.344\) to \(0.508\)
		for \(k=-1\), and from approximately \(0.228\) to \(0.323\) for
		\(k=+1\).  All three branches therefore have positive heat capacity
		throughout the range shown.
		
		Figure~\ref{fig:st-z2} presents the corresponding genuinely Lifshitz
		case.  Here the theory and the asymptotic Lifshitz data are held fixed,
		and the family is followed at fixed
		\(h_1\sqrt{r_0}=\sqrt{3}\) while \(r_0\) is varied.  The planar
		branch obeys \(T\propto r_0^2\) and
		\(\mathcal{S}\propto r_0^3\), giving the expected logarithmic slope
		\begin{equation}
			\frac{d\log T}{d\log\mathcal{S}}
			=
			\frac{2}{3}.
		\end{equation}
		The numerical fit gives \(0.666667\).  Horizon curvature produces the
		visible bending of the \(k=\pm1\) curves: the segment slopes lie
		between approximately \(0.692\) and \(1.086\) for \(k=-1\), and
		between approximately \(0.409\) and \(0.643\) for \(k=+1\).
		They nevertheless remain positive throughout the displayed range.
		As \(r_0/L\) increases, the effect of the horizon-curvature scale
		becomes progressively less important and the curved branches approach
		the planar behavior.  
		Thus the logarithmic plots provide
		a controlled local-stability diagnostic; the stronger thermodynamic
		tests based on the conserved energy, the first law, and the free energy
		are considered next.
		
		\subsection{Fixed-theory first-law and free-energy test}
		\label{subsec:first-law-test}
		
		We now perform a direct thermodynamic consistency test on a
		one-parameter family of planar \(z=2\) black holes.  In contrast to
		the coupling comparison discussed above, the gravitational theory is
		held fixed throughout this family:
		\begin{equation}
			z=2,
			\qquad
			k=0,
			\qquad
			L=1,
			\qquad
			\left(
			\hat\mu_2,\hat\mu_3,\hat\mu_4,\hat\mu_5
			\right)
			=
			\left(
			0.04,-0.001,10^{-4},-10^{-4}
			\right).
			\label{fixedTheoryThermoParameters}
		\end{equation}
		The horizon radius is varied over
		\(1.5\leq r_0/L\leq6\), while the regular-family parameter is
		transported according to
		\begin{equation}
			h_1\sqrt{r_0}
			=
			\sqrt{3}.
			\label{fixedEtaThermoFamily}
		\end{equation}
		This prescription keeps the dimensionless horizon datum fixed and
		reproduces the reference solution \(r_0=3\), \(h_1=1\).
		
		The thermodynamic results presented below refer specifically to the
		one-parameter family defined by Eq.~\eqref{fixedEtaThermoFamily},
		rather than to arbitrary independent variations of the near-horizon
		parameter \(h_1\). In particular, keeping \(h_1\sqrt{r_0}\) fixed
		does not impose the vanishing of the slowly decaying asymptotic mode,
		nor does it by itself establish that the coefficient of this mode
		remains constant as \(r_0\) varies. Although the leading Lifshitz
		boundary data are fixed, a holographic identification of the slow
		mode as an additional deformation source and the thermodynamic
		consequences of varying such a source would require a separate
		boundary analysis. The first-law test below should therefore be
		understood as a consistency check along the specified family.
		
		For the planar five-dimensional branch, the conserved radial charge
		and the horizon quantities satisfy
		\begin{equation}
			\mathcal{C}_0
			=
			16\pi G_5 T\mathcal{S}.
			\label{C0HorizonThermoIdentity}
		\end{equation}
		Using the energy normalization discussed in
		Subsec.~\ref{subsec:energy-thermo}, the energy density for \(z=2\)
		is therefore
		\begin{equation}
			\mathcal{E}
			=
			\frac{3}{80\pi G_5}\mathcal{C}_0
			=
			\frac{3}{5}T\mathcal{S}.
			\label{z2EnergySmarr}
		\end{equation}
		The corresponding Helmholtz free-energy density is
		\begin{equation}
			\mathcal{F}
			=
			\mathcal{E}-T\mathcal{S}
			=
			-\frac{1}{40\pi G_5}\mathcal{C}_0
			=
			-\frac{2}{5}T\mathcal{S}.
			\label{z2FreeEnergy}
		\end{equation}
		
		All numerical thermodynamic results below use \(G_5=1\).
		Along the family in Eq.~\eqref{fixedEtaThermoFamily}, the numerical
		solutions obey
		\begin{equation}
			T\propto r_0^2,
			\qquad
			\mathcal{S}\propto r_0^3,
			\qquad
			\mathcal{C}_0\propto r_0^5,
			\qquad
			\mathcal{E}\propto r_0^5.
			\label{z2ThermoScaling}
		\end{equation}
		At the reference point \(r_0=3\), for example, we find
		\begin{equation}
			T
			=
			3.4121946955,
			\qquad
			\mathcal{S}
			=
			6.75,
			\qquad
			\mathcal{C}_0
			=
			1157.7303851,
		\end{equation}
		and hence
		\begin{equation}
			\mathcal{E}
			=
			13.8193885169,
			\qquad
			\mathcal{F}
			=
			-9.21292567795.
		\end{equation}
		
		We have checked the conserved charge in two independent ways.
		First, \(\mathcal{C}_0\) was evaluated from the horizon identity
		\eqref{C0HorizonThermoIdentity}.  Second, for each member of the
		19-solution family, \(\mathcal{C}_0\) was evaluated directly from the
		full conserved-charge expression on the boundary-normalized numerical
		solution.  To avoid evaluating numerical derivatives away from the
		integration mesh, this second determination was performed at an actual
		interpolation-mesh point chosen near \(r=2r_0\).  Over the complete
		family, the maximum relative difference between the mesh-based and
		horizon evaluations is approximately
		\[
		4.5\times10^{-16}.
		\]

		The first law is then tested without using the Smarr relation as an
		input.  With the leading Proca source held fixed, the appropriate
		differential relation is
		\begin{equation}
			d\mathcal{E}
			=
			T\,d\mathcal{S}.
			\label{firstLawFixedProcaSource}
		\end{equation}
		No additional Maxwell-like term \(\Phi\,dQ\) appears in this
		ensemble.  The reason is that the massive vector does not possess the
		radially conserved electric flux associated with a massless Abelian
		gauge field.  For a Maxwell field one has
		\begin{equation}
			\nabla_{\mu}F^{\mu\nu}=0,
			\qquad\Longrightarrow\qquad
			\partial_r\!\left(\sqrt{-g}\,F^{rt}\right)=0,
		\end{equation}
		so that the radial electric flux defines an independent conserved
		charge.  By contrast, the Proca equation is
		\begin{equation}
			\nabla_{\mu}F^{\mu\nu}
			=
			m^2 A^\nu,
			\qquad\Longrightarrow\qquad
			\partial_r\!\left(\sqrt{-g}\,F^{rt}\right)
			=
			m^2\sqrt{-g}\,A^t,
			\label{ProcaNoGaussCharge}
		\end{equation}
		and therefore the corresponding radial flux is not conserved.
		Moreover, for the asymptotically Lifshitz solutions considered here
		the amplitude \(q\) is not an independent black-hole integration
		constant: its value is fixed algebraically by \(z\), \(L\), and the
		quasi-topological couplings through the Lifshitz background
		conditions.  Consequently, along the fixed-theory family used in the
		thermodynamic test there is no independently variable conserved
		electric charge \(Q\), and hence no conjugate \(\Phi\,dQ\) work term.
		If instead the leading boundary value of the Proca field were varied,
		one would be working in a different ensemble and an appropriate
		source--response work term could arise; that is distinct from the
		ordinary Maxwell charge--potential contribution.
		
		For the local first-law test, the energy density is obtained from the
		independently evaluated finite-radius conserved charge rather than from
		the horizon identity.  The derivatives
		\(d\mathcal{E}/dr_0\) and \(d\mathcal{S}/dr_0\) are determined locally
		from fifth-degree polynomial fits to seven neighboring members of the
		19-point family, using a centered stencil where possible and the
		corresponding shifted stencil near the endpoints.  We define the
		relative local residual by
		\begin{equation}
			\delta_{\rm FL}^{\rm(local)}
			=
			\frac{
				\left|
				d\mathcal{E}/dr_0
				-
				T\,d\mathcal{S}/dr_0
				\right|
			}{
				\max\!\left(
				\left|d\mathcal{E}/dr_0\right|,
				\left|T\,d\mathcal{S}/dr_0\right|
				\right)
			} .
		\end{equation}
		The maximum value over the complete family is
		\[
		\delta_{\rm FL}^{\rm(local)}
		\simeq
		2.7\times10^{-16}.
		\]
		 As a complementary check, we also test the integrated first law,
		 \begin{equation}
		 	\mathcal{E}(r_0)
		 	-
		 	\mathcal{E}(r_{0,\mathrm{min}})
		 	=
		 	\int_{r_{0,\mathrm{min}}}^{r_0}
		 	T(\tilde r_0)
		 	\frac{d\mathcal{S}}{d\tilde r_0}
		 	d\tilde r_0,
		 	\label{integratedFirstLaw}
		 \end{equation}
		 where \(r_{0,\mathrm{min}}=1.5L\).  The energy difference on the
		 left-hand side is obtained from the independently evaluated
		 mesh-based conserved charge.  On the right-hand side,
		 \(T(r_0)\) is represented by a cubic interpolation through the
		 19 numerically computed temperatures, while for the planar branch
		 with \(G_5=1\),
		 \[
		 \mathcal{S}=\frac{r_0^3}{4},
		 \qquad
		 \frac{d\mathcal{S}}{dr_0}=\frac{3r_0^2}{4}.
		 \]
		 For each nontrivial family member we define
		 \begin{equation}
		 	\delta_{\rm FL}^{\rm(int)}
		 	=
		 	\frac{
		 		\left|
		 		\Delta\mathcal{E}
		 		-
		 		\displaystyle\int_{r_{0,\mathrm{min}}}^{r_0}
		 		T(\tilde r_0)
		 		\frac{d\mathcal{S}}{d\tilde r_0}
		 		d\tilde r_0
		 		\right|
		 	}{
		 		\max\!\left(
		 		|\Delta\mathcal{E}|,
		 		\left|
		 		\displaystyle\int_{r_{0,\mathrm{min}}}^{r_0}
		 		T(\tilde r_0)
		 		\frac{d\mathcal{S}}{d\tilde r_0}
		 		d\tilde r_0
		 		\right|
		 		\right)
		 	},
		 \end{equation}
		 with
		 \[
		 \Delta\mathcal{E}
		 =
		 \mathcal{E}(r_0)
		 -
		 \mathcal{E}(r_{0,\mathrm{min}}).
		 \]
		 The maximum integrated residual over the nontrivial points is
		 \[
		 \delta_{\rm FL}^{\rm(int)}
		 \simeq 2.3\times10^{-16}.
		 \]
		 Thus both the local and integrated tests verify the first law within
		 the numerical accuracy of this fixed-theory family.
		
		The heat capacity along this fixed-theory planar family is
		\begin{equation}
			C
			=
			\frac{d\mathcal{E}}{dT}
			=
			\frac{3}{2}\mathcal{S}
			>
			0,
			\label{z2PlanarHeatCapacity}
		\end{equation}
		in agreement with the positive slope of the planar curve in
		Fig.~\ref{fig:st-z2}.  Moreover,
		\(\mathcal{F}=-(2/5)T\mathcal{S}<0\) throughout the family.
		With the exact Lifshitz vacuum assigned zero free energy in the
		fixed-leading-Proca-source ensemble, this makes the black-hole branch
		thermodynamically preferred relative to that reference background.
		This statement is restricted to the ensemble and reference background
		specified here; it is not intended to establish dominance over every
		possible competing saddle or disconnected black-hole branch.
		
		It is important to distinguish these thermodynamic results from
		dynamical stability. The positivity of the heat capacity establishes
		local thermodynamic stability within the family considered here, but it
		does not determine the response of the black holes to time-dependent
		gravitational or Proca perturbations. Likewise, the absence of growing
		modes in the asymptotic radial analysis used in the numerical
		construction refers to the radial falloff of static perturbations about
		the Lifshitz fixed point and should not be interpreted as a
		quasinormal-mode stability criterion. Establishing dynamical stability
		would require a separate perturbative analysis of the coupled
		gravitational and massive-vector sectors and the corresponding
		quasinormal-mode spectrum. Such an analysis is beyond the scope of the
		present work.
		
\begin{figure}[t]
			\centering
			\includegraphics[width=0.95\textwidth]{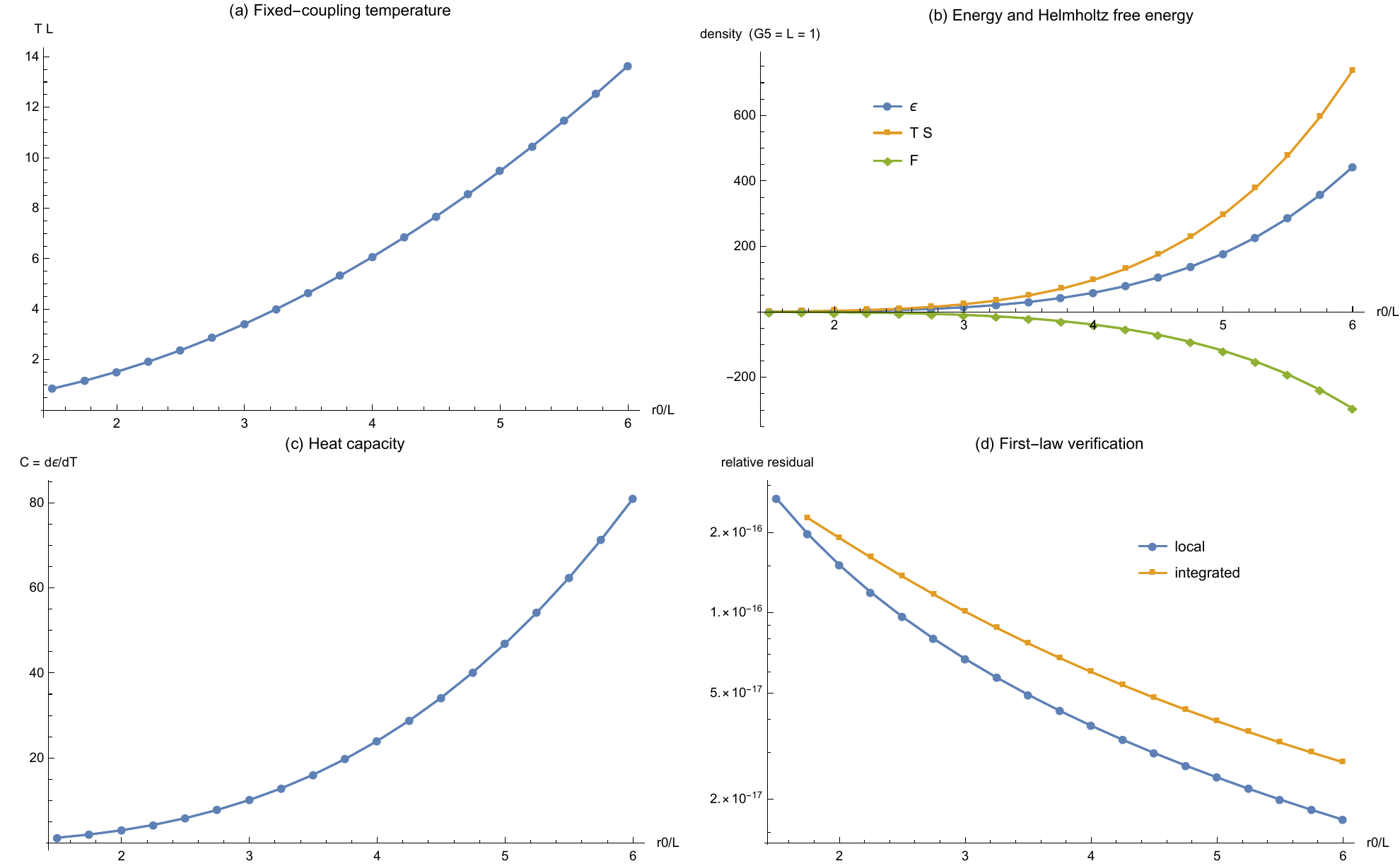}
			\caption{
				Thermodynamic consistency tests for the planar \(z=2\)
				fixed-theory family defined by
				Eqs.~\eqref{fixedTheoryThermoParameters} and
				\eqref{fixedEtaThermoFamily}, with \(G_5=L=1\).
				Panel (a) shows the dimensionless Hawking temperature \(TL\)
				as a function of \(r_0/L\).
				Panel (b) displays the energy density \(\mathcal{E}\),
				the product \(T\mathcal{S}\), and the Helmholtz
				free-energy density \(\mathcal{F}\).
				Panel (c) shows the positive heat capacity
				\(C=d\mathcal{E}/dT\).
				Panel (d) gives the relative residuals of the local and
				integrated first-law tests.  Both remain at the level of
				numerical precision throughout the family.
			}
			\label{fig:thermo-family-z2}
		\end{figure}

		Figure~\ref{fig:thermo-family-z2} summarizes these tests.  It shows
		the temperature, the energy and free-energy densities, the heat
		capacity, and both the local and integrated first-law residuals along
		the same fixed-theory family.
		
		\section{Conclusions}\label{sec:conclusion}
		
		In this paper we have investigated asymptotically Lifshitz black holes in quintic quasi-topological gravity. Within the representative quasi-topological family considered here, the
		extension from quartic to quintic order introduces an additional
		higher-curvature coupling while preserving a manageable set of reduced
		field equations for the static constant-curvature ansatz. We worked in a five-dimensional bulk and coupled gravity to a massive Abelian vector field, which is needed to sustain Lifshitz asymptotics when \(z\neq1\).
		
		We began with the quintic quasi-topological action, inserted the Lifshitz black-hole ansatz, and worked out the reduced equations that determine the metric functions \(f(r)\), \(g(r)\) and the gauge-field profile \(h(r)\). A radially conserved charge for the one-dimensional reduced system was also constructed. When \(z=1\), this charge collapses to the familiar polynomial involving all quasi-topological couplings through quintic order.
		
		We next examined the requirements for Lifshitz backgrounds to exist in the theory. Without matter, the existence of an exact Lifshitz background imposes algebraic relations linking the cosmological constant to the curvature couplings. When the massive vector field is switched on, the vector-field amplitude
		\(q\), the Proca mass, and the cosmological constant become fixed by the
		dynamical exponent and the quasi-topological couplings. The requirement that the vector-field amplitude \(q\) be real then places a corresponding constraint on the admissible region of coupling space.
		
		Given that closed-form black-hole solutions are absent in the generic quintic theory, we turned to numerical construction. Regular initial data were generated just outside the horizon from the near-horizon expansion, and the first-order system was integrated outward while enforcing the asymptotically Lifshitz boundary conditions. We produced numerical solutions for the representative cases \(z=1\)
		and \(z=2\), covering the three horizon topologies \(k=-1,0,+1\).
		The numerical formalism itself is not restricted to these two values:
		for other admissible \(z\), the Lifshitz-background parameters and
		asymptotic mode spectrum must be recomputed, and any physical growing
		mode must be removed by the appropriate shooting condition. The metric functions and gauge-field profiles share the overall structure familiar from the cubic and quartic quasi-topological Lifshitz cases: the fields vanish correctly at the horizon, develop radial profiles that depend on the topology and the chosen parameters, and relax to the Lifshitz background at large radius.
		
		A controlled comparison with the genuine quartic point
		\(\hat\mu_5=0\) makes the fifth-order contribution quantitative.
		Keeping \(n\), \(z\), \(k\), \(L\), \(r_0\),
		\(\hat\mu_2\), \(\hat\mu_3\), \(\hat\mu_4\), and \(h_1\)
		fixed, while consistently recomputing the background-constrained
		\(q^2\) and \(\Lambda\), shows that the representative
		\(\hat\mu_5=-10^{-4}\) solution is indeed extremely close to the
		quartic one. Its largest normalized profile difference is
		approximately \(9.0\times10^{-4}\), while its Hawking temperature
		differs by only \(0.0287\%\). In contrast,
		\(\hat\mu_5=-0.05\) and \(-0.10\) generate maximum normalized
		profile differences of approximately \(0.295\) and \(0.474\),
		together with temperature shifts of \(8.13\%\) and \(12.30\%\),
		respectively. Thus the near-overlap of these numerical
		profiles results from the very small quintic coupling used in those
		illustrative examples, rather than from an absence of a fifth-order
		effect.
		
		We further investigated the thermodynamics of the black holes. The
		entropy density was evaluated with the Wald prescription and the
		Hawking temperature from the near-horizon geometry. The dimensionless \(\log(TL)\)--\(\log(4G_5\mathcal{S}/L^3)\) plots
		for \(z=1\) and \(z=2\) show positive heat capacity for all three
		horizon topologies over the displayed ranges. For the planar
		\(z=2\) branch we additionally constructed a 19-member,
		fixed-theory family and used the radially conserved charge
		\(\mathcal{C}_0\) to determine the energy density. The conserved
		charge evaluated directly from the numerical solutions agrees with
		its horizon value to better than \(5\times10^{-16}\). Along this
		family the differential relation
		\(d\mathcal{E}=T\,d\mathcal{S}\) and its integrated form are both
		satisfied with maximum relative residuals of order \(10^{-16}\).
		The heat capacity is positive,
		\(C=(3/2)\mathcal{S}\), while the Helmholtz free-energy density obeys
		\(\mathcal{F}=-(2/5)T\mathcal{S}<0\). 
		These thermodynamic results refer to the specific one-parameter planar
		family defined by \(h_1\sqrt{r_0}=\sqrt{3}\), with the gravitational
		couplings and leading Lifshitz boundary data held fixed. Along this
		family, the heat capacity is positive and the Helmholtz free-energy
		density is negative relative to the adopted Lifshitz vacuum
		normalization. However, this prescription does not independently fix
		the coefficient of the slowly decaying asymptotic mode. Consequently,
		these results should not be interpreted as establishing thermodynamic
		dominance over branches with potentially different holographic
		deformation sources. A definitive comparison would require identifying
		the associated boundary source and specifying the corresponding
		thermodynamic ensemble.
		
		These thermodynamic conclusions should not be confused with a statement
		of dynamical stability. We have not analyzed time-dependent gravitational
		and Proca perturbations or computed the corresponding quasinormal-mode
		spectrum; consequently, the dynamical stability of the numerical black
		holes remains an open question.
		These results show that the quintic quasi-topological interaction
		provides a consistent higher-order extension of the cubic and
		quartic Lifshitz black holes studied previously. Within the regular
		branch examined here, the fifth-order coupling produces quantitatively
		resolved deformations of both the geometry and the Hawking temperature
		while preserving the overall qualitative structure of the lower-order
		solutions.
		
		A more exhaustive exploration of the full quintic parameter space, the dependence on physical near-horizon family parameters and
possible asymptotic boundary prescriptions, the complete phase structure including possible competing black-hole branches and saddles, and potential holographic restrictions--for instance, causality, positivity of energy flux, or quasinormal-mode stability--would all be valuable avenues for future investigation.

		\appendix
		
		\section{Curvature invariants and coupling conventions}
		\label{app:curvature-densities}

		This appendix collects the curvature-invariant basis and the coupling conventions adopted in the main body of the paper. The gravitational action includes the Einstein-Hilbert term, the Gauss-Bonnet density, and quasi-topological curvature terms through fifth order.
		The first two densities are
		\begin{equation}
			\mathcal{L}_1=R,
			\qquad
			\mathcal{L}_2
			=
			R_{abcd}R^{abcd}
			-4R_{ab}R^{ab}
			+R^2 .
		\end{equation}
		The cubic quasi-topological density is written as
		\begin{align}
			\mathcal{L}_3
			={}&
			R_a{}^c{}_b{}^d
			R_c{}^e{}_d{}^f
			R_e{}^a{}_f{}^b
			+
			\frac{1}{(2n-1)(n-3)}
			\bigg[
			\frac{3(3n-5)}{8}R_{abcd}R^{abcd}R
			\nonumber\\
			&-3(n-1)R_{abcd}R^{abc}{}_{e}R^{de}
			+3(n+1)R_{abcd}R^{ac}R^{bd}
			+6(n-1)R_a{}^bR_b{}^cR_c{}^a
			\nonumber\\
			&-\frac{3(3n-1)}{2}R_a{}^bR_b{}^aR
			+\frac{3(n+1)}{8}R^3
			\bigg].
		\end{align}
		The quartic quasi-topological density is expressed as
		\begin{align}
			\mathcal{L}_4
			={}&
			b_1R_{abcd}R^{cdef}R^{hg}{}_{ef}R_{hg}{}^{ab}
			+b_2R_{abcd}R^{abcd}R_{ef}R^{ef}
			+b_3RR_{ab}R^{ac}R_c{}^b
			\nonumber\\
			&+b_4(R_{abcd}R^{abcd})^2
			+b_5R_{ab}R^{ac}R_{cd}R^{db}
			+b_6RR_{abcd}R^{ac}R^{db}
			\nonumber\\
			&+b_7R_{abcd}R^{ac}R^{be}R^d{}_{e}
			+b_8R_{abcd}R^{acef}R^b{}_{e}R^d{}_{f}
			+b_9R_{abcd}R^{ac}R_{ef}R^{bedf}
			\nonumber\\
			&+b_{10}R^4
			+b_{11}R^2R_{abcd}R^{abcd}
			+b_{12}R^2R_{ab}R^{ab}
			\nonumber\\
			&+b_{13}R_{abcd}R^{abef}R_{ef}{}^c{}_{g}R^{dg}
			+b_{14}R_{abcd}R^{aecf}R_{gehf}R^{gbhd}.
			\label{L4Appendix}
		\end{align}
		The quintic quasi-topological density is written in the invariant basis
		\begin{align}
			\mathcal{L}_5
			={}&
			c_1RR_b{}^aR_c{}^bR_d{}^cR_a{}^d
			+c_2RR_b{}^aR_a{}^bR_{ef}{}^{cd}R_{cd}{}^{ef}
			+c_3RR_c{}^aR_d{}^bR_{ef}{}^{cd}R_{ab}{}^{ef}
			\nonumber\\
			&+c_4R_b{}^aR_a{}^bR_d{}^cR_e{}^dR_c{}^e
			+c_5R_b{}^aR_c{}^bR_a{}^cR_{fg}{}^{de}R_{de}{}^{fg}
			\nonumber\\
			&+c_6R_b{}^aR_d{}^bR_f{}^cR_{ag}{}^{de}R_{ce}{}^{fg}
			+c_7R_b{}^aR_d{}^bR_f{}^cR_{cg}{}^{de}R_{ae}{}^{fg}
			\nonumber\\
			&+c_8R_b{}^aR_c{}^bR_{ae}{}^{cd}R_{gh}{}^{ef}R_{df}{}^{gh}
			+c_9R_b{}^aR_c{}^bR_{ef}{}^{cd}R_{gh}{}^{ef}R_{ad}{}^{gh}
			\nonumber\\
			&+c_{10}R_b{}^aR_c{}^bR_{eg}{}^{cd}R_{ah}{}^{ef}R_{df}{}^{gh}
			+c_{11}R_c{}^aR_d{}^bR_{ab}{}^{cd}R_{gh}{}^{ef}R_{ef}{}^{gh}
			\nonumber\\
			&+c_{12}R_c{}^aR_d{}^bR_{ae}{}^{cd}R_{gh}{}^{ef}R_{bf}{}^{gh}
			+c_{13}R_c{}^aR_d{}^bR_{ef}{}^{cd}R_{gh}{}^{ef}R_{ab}{}^{gh}
			\nonumber\\
			&+c_{14}R_c{}^aR_d{}^bR_{eg}{}^{cd}R_{ah}{}^{ef}R_{bf}{}^{gh}
			+c_{15}R_c{}^aR_e{}^bR_{af}{}^{cd}R_{gh}{}^{ef}R_{bd}{}^{gh}
			\nonumber\\
			&+c_{16}R_b{}^aR_{ad}{}^{bc}R_{fh}{}^{de}R_{ci}{}^{fg}R_{eg}{}^{hi}
			+c_{17}R_b{}^aR_{de}{}^{bc}R_{cf}{}^{de}R_{hi}{}^{fg}R_{ag}{}^{hi}
			\nonumber\\
			&+c_{18}R_b{}^aR_{df}{}^{bc}R_{ac}{}^{de}R_{hi}{}^{fg}R_{eg}{}^{hi}
			+c_{19}R_b{}^aR_{df}{}^{bc}R_{ah}{}^{de}R_{ei}{}^{fg}R_{cg}{}^{hi}
			\nonumber\\
			&+c_{20}R_b{}^aR_{df}{}^{bc}R_{gh}{}^{de}R_{ei}{}^{fg}R_{ac}{}^{hi}
			+c_{21}R_{cd}{}^{ab}R_{eg}{}^{cd}R_{ai}{}^{ef}R_{fj}{}^{gh}R_{bh}{}^{ij}
			\nonumber\\
			&+c_{22}R_{ce}{}^{ab}R_{af}{}^{cd}R_{gi}{}^{ef}R_{bj}{}^{gh}R_{dh}{}^{ij}
			+c_{23}R_{ce}{}^{ab}R_{ag}{}^{cd}R_{bi}{}^{ef}R_{fj}{}^{gh}R_{dh}{}^{ij}
			\nonumber\\
			&+c_{24}R_{ce}{}^{ab}R_{fg}{}^{cd}R_{hi}{}^{ef}R_{aj}{}^{gh}R_{bd}{}^{ij}.
			\label{L5Appendix}
		\end{align}
		
		The coefficients \(b_i\) and \(c_i\) are fixed by demanding that, for the static constant-curvature ansatz, the reduced equations take the quasi-topological form employed in the main text. For the quartic density we adopt the standard coefficients of quartic quasi-topological gravity, while for the quintic density we use the representative quintic choice given in Ref.~\cite{CisternaGuajardoHassaineOliva2017}. With these conventions, inserting the Lifshitz ansatz into the action produces the reduced polynomial
		\begin{equation}
			-\Psi
			+\hat\mu_2\Psi^2
			+\hat\mu_3\Psi^3
			+\hat\mu_4\Psi^4
			+\hat\mu_5\Psi^5,
		\end{equation}
		in which
		\begin{equation}
			\Psi(r)=g(r)-\frac{kL^2}{r^2}.
		\end{equation}
		The dimensionless couplings \(\hat\mu_i\) that appear in the reduced equations are those used throughout the main text. The aim of this appendix is simply to state the invariant basis and the conventions behind the reduced equations; the full coefficient tables for the quartic and quintic densities are rather lengthy and are taken, respectively, from the quartic and quintic quasi-topological constructions of Refs.~\cite{DehghaniVahidinia2013,CisternaGuajardoHassaineOliva2017}.
		
		\section{Near-horizon coefficients and numerical initial data}
		\label{app:near-horizon-coefficients}
		
		This appendix collects the supplementary near-horizon data needed to
		start the numerical integration. The near-horizon expansions and the
		condition \(h_0=0\) were already presented in
		Eqs.~\eqref{near-hor} and \eqref{h0zero}. The coefficient \(g_1\) is
		determined by Eq.~\eqref{g1compact}, whereas \(f_1\) and \(h_1\)
		remain free at the level of the raw near-horizon expansion but have
		different physical roles. The coefficient \(f_1\) represents the
		residual normalization of the time coordinate, while for \(z>1\)
		\(h_1\) labels the regular physical family of solutions.
		
		The first nontrivial coefficient in the gauge-field expansion takes the form
		\begin{equation}
			h_2
			=
			-\frac{h_1}{2r_0^2g_1}
			\frac{\mathcal{N}_h}{\mathcal{D}_h},
			\label{h2appendix}
		\end{equation}
		with
		\begin{align}
			\mathcal{N}_h
			={}&
			\bigg[
			\Bigl(
			\bigl[
			1-2\hat\mu_2-3\hat\mu_3-4\hat\mu_4-5\hat\mu_5
			\bigr]h_1^2
			-2g_1
			\Bigr)z
			\nonumber\\
			&\hspace{1.2cm}
			+\bigl(
			-1+2\hat\mu_2+3\hat\mu_3+4\hat\mu_4+5\hat\mu_5
			\bigr)h_1^2
			-3g_1
			\bigg]r_0^9
			\nonumber\\
			&+3zr_0^8
			-\bigl(4z+6\bigr)\hat\mu_2L^2k g_1 r_0^7
			+6z\hat\mu_2L^2k r_0^6
			\nonumber\\
			&+\bigl(9+6z\bigr)\hat\mu_3L^4k^2 g_1 r_0^5
			-9z\hat\mu_3L^4k^2 r_0^4
			\nonumber\\
			&-\bigl(8z+12\bigr)\hat\mu_4L^6k^3 g_1 r_0^3
			+12z\hat\mu_4L^6k^3 r_0^2
			\nonumber\\
			&+\bigl(10z+15\bigr)\hat\mu_5L^8k^4 g_1 r_0
			-15z\hat\mu_5L^8k^4 ,
			\label{Nhappendix}
		\end{align}
		and
		\begin{equation}
			\mathcal{D}_h
			=
			-r_0^8
			-2\hat\mu_2L^2k r_0^6
			+3\hat\mu_3L^4k^2 r_0^4
			-4\hat\mu_4L^6k^3 r_0^2
			+5\hat\mu_5L^8k^4 .
			\label{Dhappendix}
		\end{equation}
		
	The same approach yields \(f_2\), \(g_2\), and the coefficients at all
	higher-orders. At each power of \((r-r_0)\), the field equations supply
	algebraic conditions involving the unknown expansion coefficients.
	Once \(r_0\), \(k\), \(z\), \(L\), the couplings \(\hat\mu_i\), and
	the two free coefficients \(f_1\) and \(h_1\) of the raw near-horizon
	series are specified, these algebraic conditions recursively fix every
	remaining coefficient. As discussed in the main text, \(f_1\) controls
	the residual time normalization, whereas for \(z>1\), \(h_1\) labels
	the regular physical family.
		
		In the numerical calculation the integration is not started precisely at the horizon but rather at
		\begin{equation}
			r = r_0 + \epsilon,
			\qquad
			0 < \epsilon \ll 1 .
		\end{equation}
		The initial data are generated by evaluating the near-horizon series at \(r_0+\epsilon\):
		\begin{align}
			f(r_0+\epsilon)
			&=
			f_1\Bigl[
			\epsilon + f_2\epsilon^2 + f_3\epsilon^3 + \cdots
			\Bigr],
			\nonumber\\
			g(r_0+\epsilon)
			&=
			g_1\epsilon + g_2\epsilon^2 + g_3\epsilon^3 + \cdots,
			\nonumber\\
			h(r_0+\epsilon)
			&=
			f_1^{1/2}
			\Bigl[
			h_1\epsilon + h_2\epsilon^2 + h_3\epsilon^3 + \cdots
			\Bigr],
			\nonumber\\
			j(r_0+\epsilon)
			&=
			h'(r_0+\epsilon)
			=
			f_1^{1/2}
			\Bigl[
			h_1 + 2h_2\epsilon + 3h_3\epsilon^2 + \cdots
			\Bigr].
			\label{initialdataappendix}
		\end{align}
		These initial values are then fed into the first-order system described
		in Sec.~\ref{subsec:first-order-system} and evolved outward from
		\(r=r_0+\epsilon\). As discussed in
		Sec.~\ref{subsec:asymptoticbehavior}, the physical asymptotic spectrum
		of the branches considered in this work contains no growing mode.
		Consequently, \(h_1\) is not tuned to cancel an unstable asymptotic
		mode; rather, for \(z>1\) it labels the regular family of solutions.
		The coefficient \(f_1\), on the other hand, represents the residual
		normalization of the time coordinate.
		
		After the outward integration, let
		\[
		f_\infty=\lim_{r\to\infty}f(r).
		\]
		The residual time normalization is then fixed by introducing
		\begin{equation}
			F(r)=\frac{f(r)}{f_\infty},
			\qquad
			H(r)=\frac{h(r)}{\sqrt{f_\infty}}.
		\end{equation}
		For a solution approaching the Lifshitz fixed point, the normalized
		profiles satisfy
		\begin{equation}
			\lim_{r\to\infty}F(r)=1,\qquad
			\lim_{r\to\infty}g(r)=1,\qquad
			\lim_{r\to\infty}H(r)=1.
		\end{equation}

	\end{document}